\documentclass[twocolumn, 10pt]{IEEEtran}
\IEEEoverridecommandlockouts
\usepackage{algorithm}
\usepackage{graphicx}
\usepackage[noend]{algpseudocode}
\usepackage{url} 
\usepackage{etex} 
\usepackage{enumerate}
\usepackage{tikz}
\usetikzlibrary{shapes,arrows}
\usepackage{pstricks}
\usepackage{pst-node,pst-blur}
\usetikzlibrary{arrows}
\usepackage{amsfonts}
\usepackage{tabularx}
\usepackage{subfigure}
\usepackage{amssymb}
\usepackage{steinmetz}   
\usepackage{booktabs}
\usepackage{multirow}
\usepackage{epsfig}
\usepackage{epstopdf}
\usepackage{array}

\usepackage[noadjust]{cite}

\newtheorem{remark}{Remark}

\newtheorem{proposition}{Proposition}
\newtheorem{lemma}{Lemma}

\usepackage[cmex10]{amsmath}
\usepackage[cmex10]{mathtools}

\newcommand\abs[1]{\left\lvert#1\right\rvert}
\newcommand\norm[1]{\left\lVert#1\right\rVert}

\allowdisplaybreaks  
   
 \makeatletter 
 \def\@eqnnum{{\normalsize \normalcolor (\theequation)}} 
  \makeatother
  
\begin{document}
\title{Optimal Channel Estimation for Reciprocity-Based Backscattering with a Full-Duplex MIMO Reader}
\author{Deepak Mishra,~\IEEEmembership{Member,~IEEE,} and Erik G. Larsson,~\IEEEmembership{Fellow,~IEEE}
	\thanks{D. Mishra and E. G. Larsson are with the Communication Systems Division of the Department of Electrical Engineering (ISY) at the Link\"oping University, 581 83 Link\"oping, Sweden (emails: \{deepak.mishra, erik.g.larsson\}@liu.se).}
	\thanks{This work is supported by  ELLIIT and the Swedish Research Council (VR).}
	\thanks{A preliminary five-page conference version~\cite{SPAWC18} of this work will be presented at IEEE SPAWC, Kalamata, Greece, June 2018.}}
\maketitle

\begin{abstract}
Backscatter communication (BSC) technology can enable ubiquitous deployment of low-cost sustainable wireless devices. In this work we investigate the efficacy of a full-duplex multiple-input-multiple-output (MIMO) reader for enhancing the limited communication range of monostatic BSC systems.  As this performance is strongly influenced by the channel estimation (CE) quality, we first derive a novel least-squares estimator for the forward and backward links between the reader and the tag, assuming that  reciprocity holds and $K$ orthogonal pilots are transmitted from the first $K$ antennas of an $N$ antenna reader. We also obtain the corresponding linear minimum-mean square-error estimate for the backscattered channel. After defining the transceiver design at the reader using these estimates, we jointly optimize the number of orthogonal pilots and energy allocation for the CE and information decoding phases to maximize the average backscattered signal-to-noise ratio (SNR) for efficiently decoding the tag's messages. The unimodality of this SNR in optimization variables along with a tight analytical approximation for the jointly global optimal design is also discoursed. Lastly, the selected numerical results  validate the proposed analysis, present key insights into the optimal resource utilization at reader, and quantify the achievable gains over the benchmark schemes.  
\end{abstract}



\color{black}
\section{Introduction and Background}\label{sec:intro}
\color{black}
Backscatter communication (BSC) has emerged as a promising technology that can help in practical realization of sustainable Internet of Things (IoT)~\cite{BSC-IoT,LoRa-BSC}. This technology thrives on its capability to use low-power passive devices like envelope detectors,  comparators, and impedance controllers, instead of more costly and bulkier conventional  radio frequency (RF) chain components such as  local oscillators, mixers, and converters~\cite{Mag-Amb-BSC}. 
However, the limited BSC range and low achievable bit rate are its  major fundamental bottlenecks~\cite{BSC-Cascaded}. 

\subsection{State-of-the-Art}\label{sec:rw}  
BSC systems generally comprise a power-unlimited reader and low-power tags~\cite{inv-BSC-MIMO}. As the tag does not have its own transmission circuitry, it relies on the carrier transmission from the emitter for first powering itself and then backscattering its data to the reader by appending information to the backscattered  carrier. 
So, instead of actively generating RF signals to communicate with reader, the tag simply modulates the load impedance of its antenna(s) to  reflect or absorb the received carrier signal~\cite{Coded-QAM} and thereby changing the amplitudes and phases of the backscattered signal at reader. There are three main types of BSC models as investigated in the  literature:
\begin{itemize}
    \item \textit{Monostatic}: Here, the carrier emitter and backscattered signal reader are same entities. 
    They may or may not share the antennas for concurrent carrier transmission to and backscattered signal reception from the tag, leading respectively to the full-duplex or dyadic  architectures~\cite{inv-BSC-MIMO}.   
    \item  \textit{Bi-static:}  The emitter and reader are two different entities placed geographically apart to achieve a longer range~\cite{Bistatic-BCS}.
    \item  \textit{Ambient:} Here, emitter is an uncontrollable source and the reader  decodes this     backscattered ambient signal~\cite{Mag-Amb-BSC}. 
\end{itemize}

\begin{figure}[!t]
	\centering 
	\includegraphics[width=3.48in]{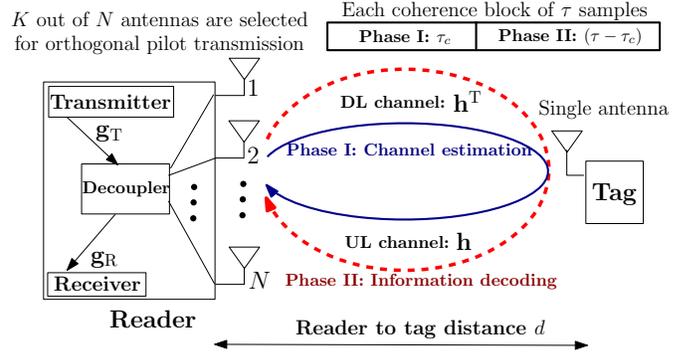}\vspace{-2mm} 
	\caption{\small Monostatic backscatter communication model with a full-duplex antenna array reader, exploiting  the proposed optimal channel estimation with orthogonal pilots transmission from  first $K$ antennas.}
	\label{fig:model} 
\end{figure} 
\textcolor{black}{As shown in Fig.~\ref{fig:model}, we consider a monostatic BSC system with a multiantenna reader working in the full-duplex mode. 
Each antenna element is used for both the unmodulated carrier emission in the downlink and  backscattered signal reception from the tag in the uplink. In contrast to full-duplex operation in conventional communication systems involving independently modulated information signals being simultaneously transmitted and received, the unmodulated carrier leakage can be much efficiently suppressed~\cite{FD-Just} in monostatic full-duplex BSC systems~\cite{FD-jnl-new}.} The adopted monostatic configuration provides the opportunity of using a large antenna array at the reader, to   maximize the BSC range while meeting the desired rate requirements.
This in turn is made possible by the   beamforming (array) gains for both transmission to and reception from the tag. However, these performance gains of  multiple-input-multiple-output (MIMO) BSC system with multiantenna reader are strongly influenced by the underlying channel estimation (CE) and tag signal detection errors. Noting that the tag-to-reader backscatter uplink  is  coupled  to  the reader-to-tag  downlink, novel higher order modulation  schemes were investigated in \cite{inv-BSC-MIMO,Coded-QAM} for the monostatic BSC systems like the Radio-frequency identification (RFID) devices. A frequency-modulated
continuous-wave based RFID system with monostatic reader, whose one antenna was dedicated for transmission and remaining for  the reception of backscattered signals, was studied in~\cite{RFID-TSP} to precisely determine the number of active tags and their positions by implementing the matrix reconstruction and stacking techniques. Further, the practical implementation of the full-duplex monostatic BSC system with single antenna Wi-Fi access point as the reader was presented in \cite{BackFi}.  

Other than these monostatic configurations, designing efficient detection techniques for recovering the messages from multiple tags due to the ambient backscattering has also gained recent  interest~\cite{FD-BSC-2ant,Amb-BCS-Ana,semi-coh-BSC,OFDM-BSC,CE-CL-BSC}. Considering a full-duplex  two antenna monostatic BSC model, authors in \cite{FD-BSC-2ant} investigated ambient backscattering from a Wi-Fi transmitter that while transmitting to its client using one antenna, uses the second antenna to simultaneously receive the backscattered signal from the tag. Assuming that the BSC channel is perfectly known at the reader, a linear minimum mean square error (LMMSE) based  estimate  of the  channel  between its transmit and receive antenna was first used to eliminate self-interference and then a maximum likelihood (ML) detector was proposed to decode the tag's messages, received due to the ambient backscattering. \textcolor{black}{Investigating blind CE algorithms for ambient BSC, authors in \cite{CE-CL-BSC} obtained the estimates for absolute  values of: (a) channel coefficient for RF source to tag link, and (b) the composite channel coefficient involving the sum of direct and backscattered  (which is the scaled product of forward and backward coefficients) channels. However, the actual complex values of the individual forward and backward channel coefficients in multiantenna BSC were not estimated.} 

\textcolor{black}{Lastly, we discuss another related field of works~\cite{Prod-Relay-CE-TSP,Prod-Relay-CE-TVT} (and references therein) that involve the estimation of product channels in the half-duplex two-way amplify-and-forward (AF) relaying networks. Other than the fact that these setups involve product or cascaded channels as in the BSC settings, there are some significant differences. First, compared to AF relays assisting in source-to-destination transmission by actively generating new information signals, BSC does not involve a transmitter module at the tag. Second, these AF relays generally~\cite{Prod-Relay-CE-TSP,Prod-Relay-CE-TVT} adopt the spectrally-inefficient half-duplex mode because the underlying severe self-interference in full-duplex implementation  needs complex interference cancellation techniques. Thirdly, the CE in AF relaying scenarios involve two-phases, where in the first phase source-to-relay channel is estimated at the relay. Then in the second phase, the cascaded source-to-relay-to-destination channel is estimated by the destination using CE outcome of the first phase as feedback sent by relay. Therefore, the existing CE algorithms developed for AF relaying networks cannot be used in BSC because tags do not have any radio resources like AF relays to help in separating out the two channels in the product.}

\color{black}\subsection{Paper Organization and Notations Used} 
After presenting the basic motivation, application scope, and the key contributions of this work in Section~\ref{sec:motiv}, the adopted system model and the proposed CE protocol in Section~\ref{sec:model}. Thereafter, the problem definition and the building blocks   for the proposed CE are outlined in Section~\ref{sec:form}. \color{black} Section~\ref{sec:est} discloses the novel solution methodology to obtain the estimate for the backscattered channel vector while minimizing the underlying least-squares (LS) error. The performance analysis for the effective average BSC SNR available for information decoding (ID) based on the optimal precoder and decoder designs is carried out in Section~\ref{sec:ana}. Both the individual and joint optimization of reader's total energy and orthogonal PC to be used during CE phase is conducted in ~\ref{sec:opt}. Section~\ref{sec:res} presents the detailed numerical investigation, with the concluding remarks being provided in Section~\ref{sec:concl}.

Throughout this paper, vectors and matrices are respectively denoted by boldface lowercase and capital letters. $\mathbf{A}^{\mathrm H}$, $\mathbf{A}^{\mathrm T}$, and $\mathbf{A}^{\mathrm *}$ respectively denote the Hermitian transpose, transpose, and conjugate of matrix $\mathbf{A}$. $\mathbf{0}_{n\times n}$ and $\mathbf{I}_{n}$ respectively represent  $n\times n$ zero and identity matrices.  $[\mathbf{A}]_{i,j}$ stands for   $(i,j)$-th element of matrix $\mathbf{A}$ and $[\mathbf{a}]_{i}$ stands for  $i$-th element of vector $\mathbf{a}$. With $\mathrm{Tr}\left(\mathbf{A}\right)$  being the trace,  $\lVert\,\cdot\,\rVert$ and $\left|\,\cdot\,\right|$ respectively represent Frobenius norm of a complex matrix and absolute value of a complex scalar. Expectation, covariance, and variance operators are respectively defined using $\mathbb{E}\left\lbrace\cdot\right\rbrace$, $\mathrm{cov}\left\lbrace\cdot\right\rbrace$, and $\mathrm{var}\left\lbrace\cdot\right\rbrace$. Lastly, with $j=\sqrt{-1}$, $\mathbb{R}$ and  $\mathbb{C}$ respectively denoting the real and complex number sets, $\mathbb{C} \mathbb{N}\left(\boldsymbol{\mu},\mathbf{C}\right)$ denotes complex Gaussian distribution with mean $\boldsymbol{\mu}$ and covariance matrix  $\mathbf{C}$. 

\color{black}
\section{Motivation and Significance}\label{sec:motiv}
Here after highlighting the research gap addressed and the scope of this work  corroborating its practical significance, we outline the key contributions made in the subsequent sections. 

\subsection{Novelty and Scope}\label{sec:scope}\color{black} 
Since the BSC does not require any signal modulation, amplification, or retransmission, the tags can be extraordinarily small and inexpensive wireless devices. Thus, they can form an integral part of the IoT technology~\cite{BSC-IoT} for realizing ubiquitous deployment of low power devices in smart city applications and advanced fifth generation (5G) networks~\cite{LoRa-BSC}. Here, in particular the BSC system with single antenna tag and multiantenna reader has gained practical importance because of two key reasons: (a) shifting the high cost and large form-factor constraints to the reader side, and (b) tag size miniaturization and cost reduction are key for numerous applications. Another, advantage of  BSC, especially the ambient one, is that it can coexist on top of existing RF-band, digital TV, and cellular communication protocols. However, the realization of all these goals is still very unrealistic  because for the monostatic BSC configurations with carrier generator and receiver sharing the same antenna(s) suffer
from the short communication range bottleneck. Further the backscattered or reflected signal quality gets severely impaired due to strong interference from other active reader in a dense deployment scenario which is also very costly. Lastly, the two-way BSC, involving cascaded channels, suffers from deeper  fades than conventional wireless channels which degrades their reliability and operational read range. 

The scope of this work includes addressing these challenges by optimally utilizing the resources at  multiantenna reader for accurate CE of backscattered link and efficiently decoding the reflected signal from tag to enable longer range  quality-of-service (QoS)-aware BSC.  \textcolor{black}{Although the optimal CE protocol presented in this work is dedicated to the monostatic BSC settings with  reciprocal tag-to-reader channel, the methodology proposed in Sections~\ref{sec:form} and~\ref{sec:est} can be extended to the nonreciprocal-monostatic or bi-static BSC systems where the  tag-to-reader and reader-to-tag channels are different. However, in contrast to the monostatic BSC where channel reciprocity can be exploited, for the ambient and bi-static settings, the CE phase needs to be divided into two subphases. In the first phase, the direct channel between the ambient source, or dedicated emitter, and reader can be estimated by keeping the tag in the \textit{silent} or  no backscattering mode~\cite{BackFi}. Thereafter, in the second phase, where the tag is in the active mode with its refection coefficient set to a pre-decided value, the estimated channel  information from the first phase can be used to separate out the estimate for the tag-to-reader channel from the product one. Detailed investigation combating practical challenges in designing an optimal CE protocol for ambient and bi-static settings is out of the current scope of this work and can be considered as an independent future study based on the outcomes of this paper. It may also be noted that, in contrast to conventional non-backscattering systems where for estimating the channel vector between an $N$-antenna  source and single-antenna receiver requires single pilot transmission, bi-static BSC with an  $N$-antenna reader and  $K$-antenna emitter will require atleast $K$ orthogonal pilots. However, for the monostatic BSC, we show later that the optimal PC for an $N$-antenna reader needs to be selected between $1$ and $N$.}

As noted from Section~\ref{sec:rw}, the existing works on multiantenna reader-based BSC either assume the availability of perfect channel state information (CSI)~\cite{BSC-Cascaded,inv-BSC-MIMO,Coded-QAM,Bistatic-BCS,FD-Just}, or focus on the detection of signals from multiple tags by using   statistical information on the ambient transmission and the BSC channel~\cite{FD-BSC-2ant,BackFi,Amb-BCS-Ana,semi-coh-BSC,OFDM-BSC}.  Focusing on the explicit   goal of optimizing the wireless energy transfer to a tag, \cite{BCS-WET-MIMO} obtained an estimate for the reader-to-tag channel by assuming that the reciprocal tag-to-reader channel is partially known, and only one reader antenna is used for reception. In contrast to these works, we present a more robust channel estimate that does not require any prior knowledge of the BSC channel. However, for those cases where prior information on channel statistics is available, we also present a LMMSE estimator (LMMSEE). Lastly, the proposed CE protocol obtains the estimates directly from the backscattered signal, without requiring any feedback from  tag.

\color{black}\subsection{Key Contributions}\label{sec:contib}\color{black}
We present, to our knowledge, the first investigation of optimal CE for the monostatic full-duplex BSC setup with an $N$ antenna reader.  As depicted in Fig.~\ref{fig:model},  the least-squares (LS) and LMMSE estimates are obtained using isotropically radiated and backscattered $K\le N$ orthogonal pilots during CE phase. Next, during the  information decoding (ID) phase,  maximum-ratio transmission (MRT) and maximum-ratio combining (MRC) 
are used along with optimal utilization of reader resources  to maximize the achievable beamforming gains.

Our specific technical contributions are summarized below. 
\begin{itemize}
	\item Joint CE and resource allocation based  optimal transmission protocol is proposed to maximize the achievable array gains during   BSC between a single antenna semi-passive tag and a monostatic full-duplex MIMO reader.
	\item For efficient CE, a novel LS estimator (LSE) for the BSC channel is derived. The global optimum of the corresponding non-linear optimization problem is computed by applying the principal eigenvector approximation to the underlying equivalent real domain transformation of the system of equations defining the solution set.
	\item From this nontrivial solution methodology, the LMMSEE for backscattered channel is also presented while accounting for the orthogonal pilot count (PC) used for CE\footnote{It may be noted that in \cite{SPAWC18} we only considered a special case of having $K=N$ while deriving the LSE, and the LMMSEE for $\mathbf{h}$ was not presented.}.
	\item A tight approximation for the average backscattered signal-to-noise ratio (SNR) available for ID is derived using the LSE or LMMSEE obtained after the CE phase involving $K$ orthogonal pilots transmission from the first $K$ antennas at reader. The concavity of this approximated SNR in the time or energy allocation for CE phase is proved along with its convexity in the integer-relaxed PC. 
	\item Using the above mentioned properties, the closed-form expression for the jointly optimal energy allocation and orthogonal PC at the reader is derived, that closely follows the globally optimal joint design maximizing the average effective backscattered SNR for carrying out ID.
	\item Numerical results are presented to validate the proposed analysis, provide optimal design insights, and quantify the achievable gains in the average BSC SNR for ID.
\end{itemize}


\section{System Model}\label{sec:model}

\subsection{Adopted BSC Channel and Tag Models}\label{sec:tag}
We consider the traditional monostatic BSC system~\cite{BCS-WET-MIMO,inv-BSC-MIMO} consisting of one multiple antenna reader, $\mathcal{R}$, with $N$ antennas,  and a  single antenna tag, $\mathcal{T}$.  To enable  full-duplex operation~\cite{FD-Just}, each of the $N$ antennas at $\mathcal{R}$ can  transmit a carrier signal to $\mathcal{T}$. Concurrently,  $\mathcal{R}$  receives the resulting backscattered signal.  This results in a composite (cascaded) multiple-input-multiple-output (MIMO) system defined by the transmission chain $\mathcal{R}$-to-$\mathcal{T}$-to-$\mathcal{R}$ (as shown in Fig.~\ref{fig:model}). \textcolor{black}{For enabling full-duplex operation, $\mathcal{R}$ includes a decoupler which comprises of automatic gain control circuits and conventional phase locked loops~\cite{FD-Just}. So, with careful adjustment of the underlying phase shifters and  attenuators, the carrier signal can be effectively suppressed out from the backscattered one at the receiver unit~\cite{FD-conf-new}. However, exploiting the fact that $\mathcal{R}$ performs an unmodulated transmission, this  decoupler can easily suppress the self-jamming carrier, while isolating the transmitter and receiver units' paths, to eventually implement the full-duplex architecture for monostatic BSC settings~\cite{FD-jnl-new}.}

We assume flat quasi-static Rayleigh block fading where the channel impulse response remains constant during a coherence interval of $\tau$ samples, and varies independently across different coherence blocks. The $\mathcal{T}$-to-$\mathcal{R}$ wireless channel is denoted by an $N\times1$ vector $\mathbf{h}\sim\mathbb{C} \mathbb{N}\left(\textbf{0}_{N\times 1},\beta\,\mathbf{I}_N\right)$. Here, parameter $\beta$ represents the average channel power gain incorporating the fading gain and propagation loss over $\mathcal{T}$-to-$\mathcal{R}$ or $\mathcal{R}$-to-$\mathcal{T}$ link.

For implementing the backscattering operation, we consider that $\mathcal{T}$ modulates the carrier received from $\mathcal{R}$ via a complex baseband signal denoted by $\mathrm{x}_{\mathcal{T}}\triangleq A-\zeta$~\cite{Bistatic-BCS}. Here, the load-independent constant $A$ is related to the antenna structure and the load-controlled reflection coefficient $\zeta\in\{\zeta_1,\zeta_2,\ldots,\zeta_V\}$ switches between $V$ distinct values to implement the desired tag modulation~\cite{BSC-IoT}. Further, we consider a semi-passive BSC system~\cite{semi-passive}, where $\mathcal{T}$ utilizing the
RF signals from $\mathcal{R}$ for backscattering, is equipped with an internal power source to support its low power on-board operations, without waiting to have enough harvested energy. This  reduces  access delay~\cite{Mag-Amb-BSC}.

\subsection{Proposed Backscattering Protocol}\label{sec:prot}
As the usage of multiple antennas at $\mathcal{R}$ can help in enabling the long range BSC by utilizing the beamforming gains, we now propose a novel backscattering protocol; see Fig.~\ref{fig:model}. Our protocol  involves estimation of the channel vector $\mathbf{h}$ from the cascaded backscattered channel matrix $\mathbf{H}\triangleq\mathbf{h}\,\mathbf{h}^{\rm T}$ when $N$ orthogonal pilots are used for CE, one from each antenna at $\mathcal{R}$. However, when considering the availability of limited number of orthogonal pilots, especially for $N\gg 1$ or multiple readers scenario, only first $K$ antennas are selected to transmit $K$ orthogonal pilots\footnote{As  the channel gains between the $N$ antenna elements at $\mathcal{R}$ and $\mathcal{T}$ are assumed to be independently and identically distributed, in general any of the $K$ antenna elements, not necessarily the first $K$ ones, can be selected.}. In this case with PC set to $K\le N$, $\mathbf{h}$ has to be estimated from   the reduced cascaded
matrix $\mathbf{H}_K\triangleq\mathbf{H}\,\mathbf{E}_K\in\mathbb{C}^{N\times K},$ where $\mathbf{E}_K\triangleq\left[\mathbf{e}_1\;\mathbf{e}_2\,\ldots\,\mathbf{e}_K\right]$ represents the $N\times K$ matrix with ones along the principal diagonal and zeros elsewhere. Here, the standard  basis  vector $\mathbf{e}_\mathrm{i}$ is an $N\times1$ column vector with a one in the $\mathrm{i}$th row, and zeros elsewhere. 

We refer to the forward channel, $\mathcal{R} $-to-$ \mathcal{T}$,   as the downlink (DL) 
and the backward channel, $\mathcal{T} $-to-$ \mathcal{R}$,   as  the uplink (UL). 
Assuming channel reciprocity~\cite{Signal-WPT-TC17,BCS-WET-MIMO}, the cascaded UL-DL channel $\mathbf{H}_K$ coefficients are  estimated   during the CE phase  from backscattered  pilot signals,    isotropically transmitted from $\mathcal{R}$. 
We divide each coherence interval of $\tau$ samples  into two phases: (i) the CE phase involving the isotropic $K$ orthogonal pilot signals transmission, and (ii) the ID phase involving MRT to $\mathcal{T}$ and MRC at $\mathcal{R}$ using the CE obtained in the
first phase.

During the CE phase of $1\le \tau_c\le \tau$ samples,  $\mathcal{R}$ transmits $K$ orthogonal pilots each of length $\tau_c$ samples from the first $K\le N$ antennas and $\mathcal{T}$ sets its refection coefficient to $\zeta_0$.  \textcolor{black}{This tag's cooperation in CE can be practically implemented as a preamble~\cite{BackFi} for each symbol transmission. Specifically, we assume that the tag does not instantaneously start its desired backscattering operation, and rather remains in a state (as characterized by $\zeta=\zeta_0$) known  to $\mathcal{R}$ during the CE phase.} The $K$ orthogonal pilots can collectively represented by a pilot signal matrix $\mathbf{S}\in\mathbb{C}^{K\times \tau_c}$. 
With $p_t$ denoting the average transmit power of $\mathcal{R}$, the orthogonal pilot signal matrix satisfies $\mathbf{S}\,\mathbf{S}^{\mathrm H}=\frac{p_t}{K}\,\tau_c\,\mathbf{I}_K$. Without  loss of generality, we assume that $\tau_c= K$, with each sample of length $L$ in seconds (so in time units, $\tau_c=KL$ seconds (s)). \textcolor{black}{Typically, as the length of samples or symbol duration in practical BSC implementations is greater $1.56\text{ micorseconds }(\mu\text{s})$~\cite[refer to ISO 18000-6C standard]{RFID-bookch}, we use $L\ge2\mu$s~\cite{BCS-WET-MIMO}.} Hence, the total energy radiated during the CE phase is denoted by $\mathrm{E}_\mathrm{c}\triangleq\norm{\mathbf{\mathbf{S}}}^2=p_t\,\tau_c.$ A  key merit of this proposed CE protocol is that all  computations occur at   $\mathcal{R}$, which has the required radio and computational resources.  


\section{Problem Definition}\label{sec:form}
Following the discussion in Section~\ref{sec:prot} and using $K$ orthogonal pilots represented by $\mathbf{S}$, the received signal matrix $\mathbf{Y}\in\mathbb{C}^{N\times K}$ at $\mathcal{R}$ during the CE phase can be written as:
\begin{eqnarray}\label{eq:rxR}
\mathbf{Y}= \mathbf{h}
\,\left(A-\zeta_0\right)\,\mathbf{h}^{\rm T}\,\mathbf{E}_K\,\mathbf{S}+\mathbf{W}=\mathbf{H}_K\,\mathbf{S}_{0}+\mathbf{W}, 
\end{eqnarray}     
where $\mathbf{S}_{0}\triangleq\left(A-\zeta_0\right)\mathbf{S}\in\mathbb{C}^{K\times K},$ and $\mathbf{W}\in\mathbb{C}^{N\times K}$ is the
complex additive white Gaussian noise (AWGN) matrix with  zero-mean independent and identically distributed entries having variance $N_0$.   We next formulate the problem of LS estimation of the BSC channel $\mathbf{h}$, based on  the received signal  $\mathbf{Y}\in\mathbb{C}^{N\times K}$. This estimate does not require any prior knowledge of the statistics of the  matrices $\mathbf{H}$ or $\mathbf{W}$. \textcolor{black}{Also, we have listed the frequently-used system parameters in Table~\ref{tab:notation}.}
\begin{table}[!t]
	\centering\color{black} 
	\caption{Description of notations used for key parameters}\label{tab:notation}\setlength\tabcolsep{3pt}
	\begin{tabular}{|c|c|} \hline   
		\textbf{Parameter}  & \textbf{Notation}\\\hline 
		Antenna elements at $\mathcal{R}$  & $N$\\\hline 
		Orthogonal PC for CE & $K$ \\\hline 
		Sample duration in s & $L$ \\\hline 
		Transmit power budget at $\mathcal{R}$ & $p_t$ \\\hline 
		Average received power at $\mathcal{T}$ & $p_r$\\\hline 
		Amplitude of tag's modulation during CE phase  & $a_0$ \\\hline 
		Average amplitude of tag's  modulation during ID phase & $\overline{a}$ \\\hline 
		AWGN variance  & $N_0$ \\\hline 
		Average channel power gain  & $\beta$ \\\hline 
		$\mathcal{R}$-to-$\mathcal{T}$ distance (or read range) & $d$ \\\hline 
		Cascaded channel matrix with PC as $K$  & $\mathbf{H}_{K}$\\\hline 
		Proposed LS-based channel estimate  & $\widehat{\mathbf{h}}_{\rm L}$\\\hline 
		Proposed LMMSE-based channel estimate & $\widehat{\mathbf{h}}_{\rm M}$\\\hline 
		Coherence block length in samples  & $\tau$ \\\hline 
		CE phase length in samples  & $\tau_c$\\\hline 
		Jointly optimal TA and PC design & $\tau_{c,\rm jo},K_{\rm jo}$\\\hline 
		Optimal TA for CE phase with PC as $K$ & $\tau_{c{\rm{a}_K}}$\\\hline 
		Effective average backscattered SNR during the ID phase  & $\overline{\gamma}$\\\hline 
		Approximation for effective average backscattered SNR $\overline{\gamma}$ & $\overline{\gamma}_{\rm a}$\\\hline 
		Average backscattered SNR during CE phase  & $\overline{\gamma}_{\rm E}$\\\hline 
		Average backscattered SNR under perfect CSI availability  & $\overline{\gamma}_{\rm id}$\\\hline 
		Average SNR threshold for optimal PC selection & $\overline{\gamma}_{\rm th}$\\\hline   
	\end{tabular}
\end{table}

\subsection{Least-Squares Optimization Formulation}\label{sec:LS-Opt}
The optimal LSE for the considered MIMO backscatter channel can be obtained by solving the following problem:
\begin{align}\nonumber
\mathcal{OP}_{\rm L}:\quad&  \underset{\mathbf{H}_K}{\text{argmin}}\;\,
\norm{\mathbf{Y}-\mathbf{H}_K\,\mathbf{S}_0}^2,\nonumber\\
&\text{subject to}\;\; ({\rm C1}):\,\mathbf{H}_K=\mathbf{h}
\,\mathbf{h}^{\rm T}\,\mathbf{E}_K. 
\end{align} 
Firstly, by ignoring the rank-one constraint $({\rm C1})$ in $\mathcal{OP}_{\rm L}$, we obtain a convex problem whose solution, denoted by $\widehat{\mathbf{H}}_{\rm L}\in\mathbb{C}^{N\times K}$, as defined in terms of the pseudo-inverse $\mathbf{S}_0^\dagger\triangleq\mathbf{S}_0^{\mathrm H}\left(\mathbf{S}_0 \mathbf{S}_0^{\mathrm H}\right)^{-1}$ of the scaled pilot matrix $\mathbf{S}_0$ \cite{kay1993fundamentals} is: 
\begin{align}\label{eq:LS-PI}
\widehat{\mathbf{H}}_{\rm L}=\mathbf{\mathbf{Y}}\;\mathbf{S}_0^{\dagger}=\frac{\mathbf{Y}\;\mathbf{S}_0^{\mathrm H}\,K}{a_0^2\,\mathrm{E}_\mathrm{c}} =\mathbf{H}_K+\frac{\mathbf{\mathbf{W}}\,\mathbf{S}_0^{\mathrm H}}{\mathrm{E}_0},
\end{align}  
where $a_0=\abs{A-\zeta_0}$ is the amplitude of modulation at $\mathcal{T}$ for the CE phase and $\mathrm{E}_0\triangleq\frac{a_0^2}{K}\,\mathrm{E}_\mathrm{c}$. Here, we have also used the fact that
$\mathbf{S}_0\,\mathbf{S}_0^{\mathrm H}=\mathrm{E}_0\,\mathbf{I}_K$.
Further, the LSE $\widehat{\mathbf{H}}_{\rm L}$  of $\mathbf{H}_K$, as  defined in \eqref{eq:LS-PI}, can be written in the following simplified form:
\begin{equation}\label{eq:LS-PI2}
\widehat{\mathbf{H}}_{\rm L}= \mathbf{H}_K+\widetilde{\mathbf{H}},
\end{equation}  
where $\widetilde{\mathbf{H}}\triangleq\frac{\mathbf{\mathbf{W}}\,\mathbf{S}_0^{\mathrm H}}{\mathrm{E}_0}$ is a linear function of $\mathbf{\mathbf{W}}$ and independent of $\mathbf{\mathbf{H}}_K$. As $\widehat{\mathbf{H}}_{\rm L}$ is a sufficient statistic for  estimating $\mathbf{H}_K$, $\mathcal{OP}_{\rm L}$ can be reformulated as an equivalent unconstrained problem $\mathcal{OP}_{\rm L1}$ defined below, by substituting the equality constraint $({\rm C1})$ in the objective and considering the identity matrix as the pilot by multiplying $\mathbf{\mathbf{Y}}$ with $\mathbf{S}_0^{\dagger}$  as defined earlier,
\begin{equation} 
\mathcal{OP}_{\rm L1}: \quad  \underset{\mathbf{h}}{\text{argmin}}\quad \Theta\left\lbrace\widehat{\mathbf{H}}_{\rm L}\right\rbrace\triangleq \norm{\widehat{\mathbf{H}}_{\rm L}-\mathbf{h}
	\,\mathbf{h}^{\rm T}\,\mathbf{E}_K}^2.  
\end{equation} 
We observe that problem $\mathcal{OP}_{\rm L1}$ is nonconvex and has multiple critical points in $\mathbf{h}$, yielding different suboptimal solutions. Also, it is worth noting that if we had $\mathbf{h}\,\mathbf{h}^{\rm H}$ with $K=N$ in the objective of $\mathcal{OP}_{\rm L1}$, instead of $\mathbf{h}\,\mathbf{h}^{\rm T}\,\mathbf{E}_K$, then a principal eigenvector based rank-one approximation for $\frac{\widehat{\mathbf{H}}_{\rm L}+\widehat{\mathbf{H}}_{\rm L}^{\rm H}}{2}$ would have yielded the desired solution. However, as the structure of $\mathcal{OP}_{\rm L1}$ is very different, in Section~\ref{sec:est} we derive the optimal solution of $\mathcal{OP}_{\rm L1}$ by first setting the derivative of the objective $\norm{\widehat{\mathbf{H}}_{\rm L}-\mathbf{h}\,\mathbf{h}^{\rm T}\,\mathbf{E}_K}^2 $ with respect to $\mathbf{h}$ equal to zero and solving it with respect to $\mathbf{h}$. We then later via an equivalent transformation to the real domain obtain a solution $\widehat{\mathbf{h}}$ (based on principal eigenvector approximation) for $\mathcal{OP}_{\rm L1}$, which although not  unique, provides the \textit{global minimum} value of the objective  $\norm{\widehat{\mathbf{H}}_{\rm L}-\mathbf{h} \mathbf{h}^{\rm T}\mathbf{E}_K}^2$ in the LS problem $\mathcal{OP}_{\rm L1}$.

\subsection{Linear Minimum Mean Squares Optimization Formulation}\label{sec:LMS}
The optimal LMMSEE for the considered MIMO BSC channel, minimizing the underlying LMMSE, can be obtained by solving the following optimization problem~\cite[eq. (4)]{LMMSE-Basic}:
\begin{eqnarray}\nonumber
\mathcal{OP}_{\rm M}:\quad  \underset{\,\mathbf{G}_0}{\text{argmin}}\quad
\mathbb{E}\left\lbrace\norm{\mathbf{G}_0\,\mathbf{Y}-\mathbf{h}\,\mathbf{h}^{\rm T}\,\mathbf{E}_K\,\mathbf{S}_0}^2\right\rbrace.
\end{eqnarray} 
For solving $\mathcal{OP}_{\rm M}$, let us first rewrite it in an alternate form by vectorizing the received signal matrix $\mathbf{Y}$ at $\mathcal{R}$ in \eqref{eq:rxR} to obtain:
\begin{equation}\label{eq:LS-PI3}
{\mathbf{y}}_{\rm v}=\mathbf{S}_{0\rm v}\,\mathbf{h}_{\rm v}+{\mathbf{w}}_{\rm v},
\end{equation}  
where ${\mathbf{y}}_{\rm v}=\mathrm{vec}\left\lbrace{\mathbf{Y}}\right\rbrace$, ${\mathbf{h}}_{\rm v}=\mathrm{vec}\left\lbrace{\mathbf{H}}\,\mathbf{E}_K\right\rbrace$, $\mathbf{S}_{0\rm v}=\mathrm{vec}\left\lbrace{\mathbf{S}}_0^{\rm T}\right\rbrace\otimes\mathbf{I}_N$, and ${\mathbf{w}}_{\rm v}=\mathrm{vec}\left\lbrace{\mathbf{W}}\right\rbrace$. So, ${\mathbf{y}}_{\rm v},{\mathbf{h}}_{\rm v},{\mathbf{w}}_{\rm v}\in\mathbb{C}^{NK\times1}$ and $\mathbf{S}_{0\rm v}\in\mathbb{C}^{NK\times NK}$. Subsequently, using these definitions, $\mathcal{OP}_{\rm M}$ can be rewritten in the following vectorized form~\cite{kay1993fundamentals}:
\begin{eqnarray}\nonumber
\mathcal{OP}_{\rm Mv}:\quad  \underset{\,\mathbf{G}}{\text{argmin}}\quad
\mathbb{E}\left\lbrace\norm{\mathbf{G}\,\mathbf{y}_{\rm v}-\mathbf{S}_{0\rm v}\,\mathbf{h}_{\rm v}}^2\right\rbrace, 
\end{eqnarray} 
whose objective on simplification can be  represented as:
\begin{align}\label{eq:tr-obj-LMMSE-1p}
\mathbb{E}\left\lbrace\norm{\mathbf{G}\,\mathbf{y}_{\rm v}-\mathbf{S}_{0\rm v}\,\mathbf{h}_{\rm v}}^2\right\rbrace =&\mathrm{Tr}\,\Big\{\left(\mathbf{G}\,\mathbf{S}_{0\rm v}-\mathbf{I}_{NK}\right)\mathbf{C}_{\mathbf{h}_{\rm v}}\big(\mathbf{S}_{0\rm v}^{\rm H}\mathbf{G}^{\rm H}  \nonumber\\
&\quad\;-\mathbf{I}_{NK}\big)+\mathbf{G}\,\mathbf{C}_{\mathbf{w}_{\rm v}}\,\mathbf{G}^{\rm H}\Big\},
\end{align}
where $\mathbf{C}_{{\mathbf{h}}_{\rm v}}\triangleq\mathbb{E}\left\lbrace\mathbf{h}_{\rm v}\,\mathbf{h}_{\rm v}^{\rm H}\right\rbrace$ and $\mathbf{C}_{\mathbf{w}_{\rm v}}\triangleq\mathbb{E}\left\lbrace\mathbf{w}\,\mathbf{w}_{\rm v}^{\rm H}\right\rbrace=N_0\,\mathbf{I}_{NK}$.

Now setting derivate of \eqref{eq:tr-obj-LMMSE-1p} with respect to $\mathbf{G}$ to zero, gives:
\begin{align}\label{eq:der-tr-obj-LMMSE-1p}
\frac{\partial\,\mathbb{E}\left\lbrace\norm{\mathbf{G}\,\mathbf{y}_{\rm v}-\mathbf{S}_{0\rm v}\,\mathbf{h}_{\rm v}}^2\right\rbrace}{\partial\mathbf{G}}=\,&\mathbf{G}^*\left(\mathbf{S}_{0\rm v}^*\mathbf{C}_{\mathbf{h}_{\rm v}}^{\rm T}\mathbf{S}_{0\rm v}^{\rm T}+N_0\,\mathbf{I}_{NK}\right)\nonumber\\
\,&-\mathbf{C}_{\mathbf{h}_{\rm v}}^{\rm T}\mathbf{S}_{0\rm v}^{\rm T}=\mathbf{0}_{{NK}\times {NK}}.
\end{align}
Solving above in $\mathbf{G}\in\mathbb{C}^{NK\times NK}$ yields the desired result as:
\begin{eqnarray} 
\mathbf{G}_{\rm opt}\triangleq\mathbf{C}_{\mathbf{h}_{\rm v}}\mathbf{S}_{0\rm v}^{\rm H}\left[\mathbf{S}_{0\rm v}\mathbf{C}_{\mathbf{h}_{\rm v}}\mathbf{S}_{0\rm v}^{\rm H}+N_0\,\mathbf{I}_{NK}\right]^{-1}.
\end{eqnarray} 

With this $\mathbf{G}_{\rm opt}$ denoting the optimal solution of $\mathcal{OP}_{\rm Mv}$, the LMMSEE $\widehat{\mathbf{H}}_{\rm M}\in\mathbb{C}^{N\times K}$ for BSC channel matrix  $\mathbf{H}_K$ as obtained from the received signal ${\mathbf{y}}_{\rm v}$ along with the availability of prior  statistical information on $\mathbf{C}_{{\mathbf{h}}_{\rm v}}$ can be obtained as:
\begin{align}\label{eq:hLMMSE}
\widehat{\mathbf{h}}_{\rm v_M}\triangleq\;&\mathrm{vec}\left\lbrace\widehat{\mathbf{H}}_{\rm M}\right\rbrace=\mathbf{G}_{\rm opt}\,{\mathbf{y}}_{\rm v},\nonumber\\
=\;&\mathbf{C}_{\mathbf{h}_{\rm v}}\mathbf{S}_{0\rm v}^{\rm H}\left[\mathbf{S}_{0\rm v}\mathbf{C}_{\mathbf{h}_{\rm v}}\mathbf{S}_{0\rm v}^{\rm H}+N_0\,\mathbf{I}_{NK}\right]^{-1} {\mathbf{y}}_{\rm v}.
\end{align} 
Using this LMMSE minimization based sufficient statistic $\widehat{\mathbf{H}}_{\rm M}$, defined in \eqref{eq:hLMMSE}, for
estimating $\mathbf{H}_K$ and following the discussion with regard to $\mathcal{OP_{\rm L1}}$ in  Section~\ref{sec:LS-Opt}, $\mathcal{OP}_{\rm M}$ can be reformulated as an equivalent problem $\mathcal{OP_{\rm M1}}$ given below,
\begin{equation} 
\mathcal{OP_{\rm M1}}: \;  \underset{\mathbf{h}}{\text{argmin}}\quad \Theta\left\lbrace\widehat{\mathbf{H}}_{\rm M}\right\rbrace= \norm{\widehat{\mathbf{H}}_{\rm M}-\mathbf{h}	\,\mathbf{h}^{\rm T}\,\mathbf{E}_K}^2.  
\end{equation} 
So like  $\mathcal{OP}_{\rm L1}$,  $\mathcal{OP}_{\rm M1}$ also involves minimizing the function $\Theta\left\lbrace\cdot\right\rbrace$ over the optimization variable $\mathbf{h}$. Hence, the solution of both $\mathcal{OP}_{\rm L1}$ and  $\mathcal{OP}_{\rm M1}$ can be obtained using same proposed novel solution methodology as outlined in the next section.


\section{Proposed Backscatter Channel Estimation}\label{sec:est} 
In this section we present a novel approach to obtain the global minimizer of the LS problems, as defined by $\mathcal{OP_{\rm L1}}$ and $\mathcal{OP_{\rm M1}}$, to respectively obtain the desired LSE and LMMSEE for the BSC channel vector $\mathbf{h}$ using $K$ orthogonal pilots during the CE phase. After that we discuss two special cases,  where either single pilot (i.e., $K=1$) from the first antenna at $\mathcal{R}$ is used, or $K=N$ orthogonal pilots are transmitted via $N$ antennas at $\mathcal{R}$. \textcolor{black}{These two special cases, for whom the estimates are obtained easily on substituting their respective $K$ values in the generic estimates as derived in Section~\ref{sec:Kpilot-CE}, exhibit very simple structures and have been later shown to be the only two possible candidates for optimal PC in Section~\ref{sec:opt}.}

\subsection{Using $K$ orthogonal Pilots for LS Channel Estimation}\label{sec:Kpilot-CE}
Following the discussions  in Sections~\ref{sec:LS-Opt} and~\ref{sec:LMS}, we can rewrite $\mathcal{OP_{\rm L1}}$ and $\mathcal{OP_{\rm M1}}$ combinedly as:
\begin{equation}\nonumber
\mathcal{OP}_K\!: \underset{\mathbf{h}}{\text{argmin}}\;\,\Theta\!\left\lbrace\widehat{\mathbf{H}}\right\rbrace\!=\! \begin{cases}
\norm{\widehat{\mathbf{H}}_{\rm L}-\mathbf{h}	\,\mathbf{h}^{\rm T}\,\mathbf{E}_K}^2, & \!\text{LSE,}\\
\norm{\widehat{\mathbf{H}}_{\rm M}-\mathbf{h}	\,\mathbf{h}^{\rm T}\,\mathbf{E}_K}^2, & \!\text{LMMSEE.}
\end{cases}
\end{equation}  

\subsubsection{\color{black}Characterizing the Critical Points} The objective of $\mathcal{OP}_K$ is to obtain $\mathbf{h}$ which minimizes the LS error $\Theta\left\lbrace\widehat{\mathbf{H}}\right\rbrace$, where $\widehat{\mathbf{H}}=\widehat{\mathbf{H}}_{\rm L}$ as defined in \eqref{eq:LS-PI2} for obtaining the LSE and $\widehat{\mathbf{H}}=\widehat{\mathbf{H}}_{\rm M}$ as defined by \eqref{eq:hLMMSE} for obtaining the LMMSEE of $\mathbf{h}$. Hence, to solve $\mathcal{OP}_K$ next we first characterize all the critical points of $\Theta\left\lbrace\widehat{\mathbf{H}}\right\rbrace$ with respect to $\mathbf{h}$, i.e., obtain all the solutions of $\frac{\partial\,\Theta\left\lbrace\widehat{\mathbf{H}}\right\rbrace}{\partial\mathbf{h}}=0$ in vector $\mathbf{h}$. 

First let us rewrite $\Theta\left\lbrace\widehat{\mathbf{H}}\right\rbrace$ in the following expanded form.
\begin{align}\label{eq:tr-obj-KP}
\norm{\widehat{\mathbf{H}}-\mathbf{h}
	\mathbf{h}^{\rm T}\mathbf{E}_K}^2 &=\mathrm{Tr}\left\lbrace \widehat{\mathbf{H}} \widehat{\mathbf{H}}^{\rm H}-\widehat{\mathbf{H}}\,\mathbf{E}_K^{\rm T}\,\mathbf{h}^*
\mathbf{h}^{\rm H} -\mathbf{h}
\mathbf{h}^{\rm T}\nonumber\right.\\
&\,\quad\left.\times\mathbf{E}_K\widehat{\mathbf{H}}^{\rm H}+\mathbf{h}
\mathbf{h}^{\rm T}\mathbf{E}_K \mathbf{E}_K^{\rm T}\mathbf{h}^*
\mathbf{h}^{\rm H}\right\rbrace.
\end{align}
Now, taking the derivate of \eqref{eq:tr-obj-KP} with respect to $\mathbf{h}$, using the rules  in~\cite[Chs. 3, 4]{complex-matrixbook} and setting it to zero, gives:
\begin{align}\label{eq:der-tr-obj-PK}
&\frac{\partial}{\partial\mathbf{h}}\norm{\widehat{\mathbf{H}}-\mathbf{h}\,\mathbf{h}^{\rm T}\mathbf{E}_K}^2=-\mathbf{h}^{\rm T}\left(\mathbf{E}_K\widehat{\mathbf{H}}^{\rm H}+\left(\mathbf{E}_K\widehat{\mathbf{H}}^{\rm H}\right)^{\rm T}\right)\nonumber\\
&\quad+ \mathbf{h}^{\rm T}\left(\mathbf{h}^*\,\mathbf{h}^{\rm H}\,\mathbf{E}_K\,\mathbf{E}_K^{\rm T}+ \mathbf{E}_K\,\mathbf{E}_K^{\rm T}\,\mathbf{h}^*\,\mathbf{h}^{\rm H}\right)=\mathbf{0}_{1\times N},
\end{align}

After applying some simplifications to \eqref{eq:der-tr-obj-PK} we obtain:
\begin{equation}\label{eq:SysEqK}
\overline{\mathbf{H}}_{\mathrm{E}}\,\mathbf{h}=\left(\mathbf{h}^*\,\mathbf{h}^{\rm H}\,\mathbf{E}_K\,\mathbf{E}_K^{\rm T}+ \mathbf{E}_K\,\mathbf{E}_K^{\rm T}\,\mathbf{h}^*\,\mathbf{h}^{\rm H}\right)\mathbf{h}
\end{equation} 
where the symmetric matrix $\overline{\mathbf{H}}_{\mathrm{E}}\in\mathbb{C}^{N\times N}$ is defined below:
\begin{equation}\label{eq:HEbar}
\overline{\mathbf{H}}_{\mathrm{E}}\triangleq\mathrm{sym}\left\lbrace\widehat{\mathbf{H}}^*\mathbf{E}_K^{\rm T}\right\rbrace=\left(\widehat{\mathbf{H}}^*\mathbf{E}_K^{\rm T}\right)^{\rm T}+\widehat{\mathbf{H}}^*\mathbf{E}_K^{\rm T}.
\end{equation} 
We can notice that \eqref{eq:SysEqK} involves solving a system of $N$ complex nonlinear equations in $N$ complex entries of $\mathbf{h}$, which is computationally very expensive if the antenna array at $\mathcal{R}$ is large ($N\gg 1$). \textcolor{black}{Therefore, we next present an alternative \textit{real domain} representation for \eqref{eq:SysEqK} that can be efficiently solved}.

\subsubsection{\color{black}Equivalent Real Domain Transformation}
With CE protocol involving transmission of $K$ orthogonal pilots  from the first $K$ antennas at $\mathcal{R}$, let us denote the first $K$ entries of $\mathbf{h}\in\mathbb{C}^{N\times 1}$ by a $K\times 1$ column vector $\mathbf{h}_K\triangleq\left[[\mathbf{h}]_1\;[\mathbf{h}]_2\,\ldots\,[\mathbf{h}]_K\right]^{\rm T}$ and the remaining $N-K$ entries by a $(N-K)\times 1$ column vector $\mathbf{h}_{\bar{K}}\triangleq\left[[\mathbf{h}]_{K+1}\;[\mathbf{h}]_{K+2}\,\ldots\,[\mathbf{h}]_N\right]^{\rm T}$. Hence,
$\mathbf{h}_{RI_K}\triangleq\left[\begin{array}{cc}
\mathrm{Re}\{\mathbf{h}_K\} \\
\mathrm{Im}\{\mathbf{h}_K\} \end{array}\right]\in\mathbb{R}^{2K\times1}$ and $\mathbf{h}_{RI_{\bar{K}}}\triangleq\left[\begin{array}{cc}
\mathrm{Re}\{\mathbf{h}_{\bar{K}}\} \\
\mathrm{Im}\{\mathbf{h}_{\bar{K}}\} \end{array}\right]\in\mathbb{R}^{2(N-K)\times1}$ represent the corresponding real vectors.
Next, letting  the real matrices $\mathrm{Re}\{\overline{\mathbf{H}}_{\mathrm{E}}\}$
and $\mathrm{Im}\{\overline{\mathbf{H}}_{\mathrm{E}}\}$  denote the real and imaginary parts of $\overline{\mathbf{H}}_{\mathrm{E}}$ defined in \eqref{eq:HEbar}, the system of $N$ nonlinear complex equations in \eqref{eq:SysEqK} is   equivalent to the following  system of $2N$ nonlinear \textit{real} equations:
\begin{align}\label{eq:real-KP}
\mathbf{Z}_{\mathrm{E}}\;\mathbf{h}_{RI}=\left[
\begin{array}{cc}
\mathbf{D} &  \mathbf{0}_{2K\times(2N-2K)}\\
\mathbf{0}_{(2N-2K)\times2K}  & \mathbf{D}
\end{array}
\right] \,\mathbf{h}_{RI},
\end{align}
where $\mathbf{Z}_{\mathrm{E}}\in\mathbb{R}^{2N\times2N}$ is a real symmetric matrix defined as:
\begin{align}\label{eq:Phi}
\mathbf{Z}_{\mathrm{E}}=\Phi\left\lbrace\overline{\mathbf{H}}_{\mathrm{E}}\right\rbrace\triangleq\left[\begin{array}{cc}
\mathrm{Re}\{\overline{\mathbf{H}}_{\mathrm{E}}\} & -\mathrm{Im}\{\overline{\mathbf{H}}_{\mathrm{E}}\} \\
-\mathrm{Im}\{\overline{\mathbf{H}}_{\mathrm{E}}\} &  -\mathrm{Re}\{\overline{\mathbf{H}}_{\mathrm{E}}\} \end{array} \right].
\end{align}
Further in \eqref{eq:real-KP},  $\mathbf{h}_{RI}\triangleq\left[\begin{array}{cc}
\mathrm{Re}\{\mathbf{h}\} \\
\mathrm{Im}\{\mathbf{h}\} \end{array}\right]\in\mathbb{R}^{2N\times1}$ is a real vector and the real diagonal matrix $\mathbf{D}\in\mathbb{R}^{N\times N}$ is defined below:
\begin{eqnarray} 
\mathbf{D}\triangleq\left[\begin{array}{cc}
\big(\norm{\mathbf{h}_K}^2+\norm{\mathbf{h}}^2\big) \mathbf{I}_{K} &  \mathbf{0}_{K\times(N-K)}\\
\mathbf{0}_{(N-K)\times K}  & \norm{\mathbf{h}_K}^2 \mathbf{I}_{N-K}  \end{array}\right].
\end{eqnarray}

Now we try to simplify this transformed real domain problem \eqref{eq:real-KP} by introducing some intermediate variables. Let  $\overline{\mathbf{H}}_{\mathrm{E}_{K}}\in\mathbb{C}^{K\times K}$ denote the submatrix obtained from the matrix $\overline{\mathbf{H}}_{\mathrm{E}}$ by choosing its first $K$ rows and first $K$ columns. Similarly, the last $N-K$ rows of $\widehat{\mathbf{H}}\in\mathbb{C}^{N\times K}$ are denoted by a $(N-K)\times K$ matrix as denoted by $\widehat{\mathbf{H}}_{\bar{K}}$ defined below:
\begin{eqnarray} 
\widehat{\mathbf{H}}_{\bar{K}}\triangleq\left[
\begin{array}{cccc}
\big[\widehat{\mathbf{H}}\big]_{K+1,1} &  [\widehat{\mathbf{H}}]_{K+1,2} & \cdots & [\widehat{\mathbf{H}}]_{K+1,K}  \\
\big[\widehat{\mathbf{H}}\big]_{K+2,1}  & \big[\widehat{\mathbf{H}}\big]_{K+2,2} & \cdots  & \big[\widehat{\mathbf{H}}\big]_{K+2,K}\\
\vdots  & \vdots & \ddots  & \vdots\\
\big[\widehat{\mathbf{H}}\big]_{N,1}  & 
\big[\widehat{\mathbf{H}}\big]_{N,2} & \cdots  & \big[\widehat{\mathbf{H}}\big]_{N,K}
\end{array}
\right].
\end{eqnarray}
Using these definitions for $\overline{\mathbf{H}}_{\mathrm{E}_{K}} $ and $\widehat{\mathbf{H}}_{\bar{K}}$, $\overline{\mathbf{H}}_{\mathrm{E}}$ in \eqref{eq:HEbar} can be equivalently  represented in a more compact form as:
\begin{eqnarray} 
\overline{\mathbf{H}}_{\mathrm{E}}=\left[\begin{array}{cc}
\overline{\mathbf{H}}_{\mathrm{E}_{K}} &  \widehat{\mathbf{H}}_{\bar{K}}^{\rm T}\\
\widehat{\mathbf{H}}_{\bar{K}}  & \mathbf{0}_{(N-K)\times(N-K)}  \end{array}\right],
\end{eqnarray}
which on substituting in \eqref{eq:real-KP}, yields an alternate system of $2N$ equations as defined below by~\eqref{eq:real-KP2}, which then needs to be solved for obtaining the solution of the LS problem $\mathcal{OP}_K$:
\begin{align}\label{eq:real-KP2}
&\left[\begin{array}{cc}
\Phi\left\lbrace\overline{\mathbf{H}}_{\mathrm{E}_{K}}\right\rbrace & \Phi\left\lbrace\widehat{\mathbf{H}}_{\bar{K}}^{\rm T}\right\rbrace\\
\Phi\left\lbrace\widehat{\mathbf{H}}_{\bar{K}}\right\rbrace &  \mathbf{0}_{2(N-K)}
\end{array}\right] 
\left[\begin{array}{cc}
\mathbf{h}_{RI_K} \\
\mathbf{h}_{RI_{\bar{K}}} \end{array}\right]=\nonumber\\
&\left[\!\!\begin{array}{cc}
\big(\norm{\mathbf{h}_K}^2+\norm{\mathbf{h}}^2\big) \mathbf{I}_{2K} &  \mathbf{0}_{2K\times2(N-K)}\\
\mathbf{0}_{2(N-K)\times2K}  & \norm{\mathbf{h}_K}^2 \mathbf{I}_{2(N-K)}  \end{array}\!\!\right] \left[\!\!\begin{array}{cc}
\mathbf{h}_{RI_K} \\
\mathbf{h}_{RI_{\bar{K}}} \end{array}\!\!\right].
\end{align}

On further simplifying \eqref{eq:real-KP2}, it can be deduced to the following system of two real nonlinear equations:
\begin{subequations}\label{eq:RemK-LS}
	\begin{align}\label{eq:RemK-LS1}
	\big(\norm{\mathbf{h}_K}^2+\norm{\mathbf{h}}^2\big) \mathbf{h}_{RI_K} = 
	\mathbf{Z}_{\mathrm{A}_K}\,
	\mathbf{h}_{RI_K}
	+ \mathbf{Z}_{\mathrm{B}_K}^{\rm T}\,\mathbf{h}_{RI_{\bar{K}}},
	\end{align} 
	\begin{align}\label{eq:RemK-LS2}
	\mathbf{Z}_{\mathrm{B}_K}\,
	\mathbf{h}_{RI_K}= \norm{\mathbf{h}_K}^2 
	\mathbf{h}_{RI_{\bar{K}}},
	\end{align}
\end{subequations}
where  $\mathbf{Z}_{\mathrm{A}_K}\triangleq\Phi\left\lbrace\overline{\mathbf{H}}_{\mathrm{E}_{K}}\right\rbrace\in\mathbb{R}^{2K\times2K}$ and $\mathbf{Z}_{\mathrm{B}_K}\triangleq\Phi\left\lbrace\widehat{\mathbf{H}}_{\bar{K}}\right\rbrace\in\mathbb{R}^{2(N-K)\times 2K}$. Here $\Phi\left\lbrace\cdot\right\rbrace$ is the complex-to-real transformation map as defined in \eqref{eq:Phi} . After simplifying \eqref{eq:RemK-LS2}, it yields:
\begin{align}\label{eq:RemK-LSb}
\mathbf{h}_{RI_{\bar{K}}}\triangleq\left[\begin{array}{cc}
\mathrm{Re}\{\mathbf{h}_{\bar{K}}\}\\
\mathrm{Im}\{\mathbf{h}_{\bar{K}}\} \end{array}\right]=\frac{1}{\norm{\mathbf{h}_K}^2 }\;\mathbf{Z}_{\mathrm{B}_K}\,\mathbf{h}_{RI_K}.
\end{align}
Finally using another deduction, as defined below, from \eqref{eq:real-KP2}:
\begin{align}\label{eq:real-KP3}
\mathbf{Z}_{\mathrm{B}_K}^{\rm T} \,\mathbf{h}_{RI_{\bar{K}}} =\big(\norm{\mathbf{h}}^2-\norm{\mathbf{h}_K}^2\big) \mathbf{h}_{RI_K},
\end{align}
in \eqref{eq:RemK-LS1}, and simplifying we obtain the following key result:
\begin{align}\label{eq:real-KP4}
\left(\!\norm{\mathbf{h}_K}^2+\norm{\mathbf{h}}^2\!\right)\! \mathbf{h}_{RI_K}\!\! \stackrel{(r0)}{=}&\, 
\mathbf{Z}_{\mathrm{A}_K}
\mathbf{h}_{RI_K}
\!+\! \left(\!\norm{\mathbf{h}}^2-\norm{\mathbf{h}_K}^2\!\right) \!\mathbf{h}_{RI_K}\nonumber\\
\mathbf{Z}_{\mathrm{A}_K}\,
\mathbf{h}_{RI_K} {=}&\,  2\norm{\mathbf{h}_K}^2\,\mathbf{h}_{RI_K},
\end{align}
where \eqref{eq:real-KP4} is written after applying rearrangements to $(r0)$.

\subsubsection{\color{black}Semi-Closed-Form Expressions for Channel Estimates} 
As \eqref{eq:real-KP4} possesses a conventional eigenvalue problem form, the solution to \eqref{eq:real-KP4} in $\mathbf{h}_{RI_K}$ is either given by a zero vector $\mathbf{h}_{RI_K}=\mathbf{0}_{2K\times1},$ or by the eigenvector corresponding to the positive eigenvalue $\norm{\mathbf{h}_K}^2$ of the matrix $\mathbf{Z}_{\mathrm{A}_K}$. Further, since $\mathcal{OP}_K$ involves minimization of $\norm{\widehat{\mathbf{H}}-\mathbf{h}
\mathbf{h}^{\rm T}\mathbf{E}_K}^2$, its global minimum value is attained at   $\mathbf{h}=\widehat{\mathbf{h}}\triangleq\mathrm{Re}\{\widehat{\mathbf{h}}\}+j\,\mathrm{Im}\{\widehat{\mathbf{h}}\}\in\mathbb{C}^{N\times1}$, whose real and imaginary components for the first $K$ entries as obtained using the maximum eigenvalue  $\lambda_{\mathrm{Z}_{K_1}}$ of $\mathbf{Z}_{\mathrm{A}_K}$ are defined in \eqref{eq:h-GLSE-KPa}. Next on substituting \eqref{eq:h-GLSE-KPa} in \eqref{eq:RemK-LS2}, the remaining $N-K$ entries of vector $\widehat{\mathbf{h}}$ are defined in \eqref{eq:h-GLSE-KPb}:
\begin{subequations}\label{eq:hhatK}
	\begin{eqnarray}\label{eq:h-GLSE-KPa}
	\left[\!\!\begin{array}{cc}
	\mathrm{Re}\{\big[\widehat{\mathbf{h}}\big]_i\}\\
	\mathrm{Im}\{\big[\widehat{\mathbf{h}}\big]_i\} \end{array}\!\!\right]\triangleq\pm\sqrt{\frac{\lambda_{{\mathrm{Z}_{K_1}}}}{2}}\,\frac{\left[\mathbf{v}_{{\mathrm{Z}_{K_1}}}\right]_i}{\norm{\mathbf{v}_{{\mathrm{Z}_{K_1}}}}},\forall i= 1,2,\cdots,K,
	\end{eqnarray}
	\begin{align}\label{eq:h-GLSE-KPb}
	\left[\!\!\begin{array}{cc}
	\mathrm{Re}\{\big[\widehat{\mathbf{h}}\big]_m\} \\
	\mathrm{Im}\{\big[\widehat{\mathbf{h}}\big]_m\} \end{array}\!\!\right]\triangleq\frac{1}{\sum\limits_{i=1}^{K}\abs{\big[\widehat{\mathbf{h}}\big]_i}^2 }&\Bigg[\sum_{i=1}^{K}\big[\mathbf{Z}_{\mathrm{B}_K}\big]_{mi}\,\mathrm{Re}\{\big[\widehat{\mathbf{h}}\big]_i\}+\nonumber\\
	&\sum_{i=K+1}^{2K}\big[\mathbf{Z}_{\mathrm{B}_K}\big]_{mi}\,\mathrm{Im}\{\big[\widehat{\mathbf{h}}\big]_{i-K}\}\Bigg],\nonumber\\
	&\,\forall m= K+1,\cdots,N. 
	\end{align}
\end{subequations} 
Here  $\mathbf{v}_{{\mathrm{Z}_{K_1}}}\in\mathbb{C}^{K\times1}$ represents the  eigenvector corresponding to the maximum eigenvalue $\lambda_{{\mathrm{Z}_{K_1}}}$ of $\mathbf{Z}_{\mathrm{A}_K}$. Further, as the sign   cancels in the product definition $\mathbf{h}\,\mathbf{h}^{\rm T}$ used in the objective of $\mathcal{OP}_K$, this LSE $\widehat{\mathbf{h}}$ yielding the global minimizer involves an unresolvable phase ambiguity and hence, is  not unique. Without loss of generality we have considered `$+$' sign for $\widehat{\mathbf{h}}$ in \eqref{eq:h-GLSE-KPa}. Moreover, as noted from the definitions for  $\mathrm{sym}\left\lbrace\cdot\right\rbrace$ and $\Phi\left\lbrace\cdot\right\rbrace$ in \eqref{eq:HEbar}  and \eqref{eq:Phi}, respectively, along with the results in \eqref{eq:real-KP} and \eqref{eq:real-KP2}, the estimate $\widehat{\mathbf{h}}$ is actually a function of $\mathbf{Z}_{\mathrm{E}}=\Phi\left\lbrace\overline{\mathbf{H}}_{\mathrm{E}}\right\rbrace=\Phi\left\lbrace\mathrm{sym}\left\lbrace\widehat{\mathbf{H}}^*\mathbf{E}_K^{\rm T}\right\rbrace\right\rbrace$. Henceforth, they can be alternatively  represented by a relationship: $\widehat{\mathbf{h}}=\Psi\left\lbrace\mathbf{Z}_{\mathrm{E}}\right\rbrace$, as defined by \eqref{eq:hhatK}. So, we can summarize that the proposed estimate for the BSC channel vector $\mathbf{h}$ based on LS error or LMMSE minimization is denoted by:
\begin{equation}\label{eq:h_CE_Final}
\widehat{\mathbf{h}}\triangleq  \begin{cases}
\Psi\left\lbrace\Phi\left\lbrace\mathrm{sym}\left\lbrace\widehat{\mathbf{H}}_{\rm L}^*\,\mathbf{E}_K^{\rm T}\right\rbrace\right\rbrace\right\rbrace, & \text{LSE,}\\
\Psi\left\lbrace\Phi\left\lbrace\mathrm{sym}\left\lbrace\widehat{\mathbf{H}}_{\rm M}^*\,\mathbf{E}_K^{\rm T}\right\rbrace\right\rbrace\right\rbrace, & \text{LMMSEE.}
\end{cases}
\end{equation}  

\textcolor{black}{Notice that although we have not resolved the phase ambiguity in the estimates, defined by \eqref{eq:h_CE_Final}, for $\mathbf{h}$, later in Section~\ref{sec:valid} we have numerically verified that under favorable channel conditions this impact of phasor mismatch between $\mathbf{h}$ and $\widehat{\mathbf{h}}$ can be practically ignored. Furthermore, a smart selection of CE time and PC also plays a significant role in combating the negative impact of this phase ambiguity, as demonstrated later in Section~\ref{sec:insights}.  Lastly, the practical significance of these derived estimates in \eqref{eq:h_CE_Final} stems from the fact that after all the complex computations and  nontrivial transformations, we have finally reduced the whole CE process to a simple semi-closed-form expression involving just  an eigen-decomposition of a $2K\times2K$ square matrix $\mathbf{Z}_{\mathrm{A}_K}$.} 

\subsection{Special Cases for PC during CE: $K=1$ or $K=N$}\label{sec:sepcial-cases}
Now we derive the estimates $\widehat{\mathbf{h}}$ for the single pilot and $K=N$ (full) pilot cases, which are shown later in Section~\ref{sec:opt} to be the only two possible candidates for the optimal PC $K$.

\subsubsection{Single Pilot Based Channel Estimation}\label{sec:1P} For $K=1$, solving \eqref{eq:real-KP4} reduces to solving the following two equations:
\begin{subequations}\label{eq:real-KP41} 
\begin{align}\label{eq:real-KP41a}
\mathrm{Re}\{\left[\overline{\mathbf{H}}_{\mathrm{E}}\right]_{11}\}&\,\mathrm{Re}\{\big[\widehat{\mathbf{h}}\big]_1\} -\mathrm{Im}\{\left[\overline{\mathbf{H}}_{\mathrm{E}}\right]_{11}\}\, \mathrm{Im}\{\big[\widehat{\mathbf{h}}\big]_1\} \nonumber\\
=& \, 2\left(\left(\mathrm{Re}\{x\}\right)^2+\left(\mathrm{Im}\{x\}\right)^2\right)\,\mathrm{Re}\{\big[\widehat{\mathbf{h}}\big]_1\},
\end{align}
\begin{align}\label{eq:real-KP41b}
-\mathrm{Im}\{\left[\overline{\mathbf{H}}_{\mathrm{E}}\right]_{11}\}&\,\mathrm{Re}\{\big[\widehat{\mathbf{h}}\big]_1\} -\mathrm{Re}\{\left[\overline{\mathbf{H}}_{\mathrm{E}}\right]_{11}\}\, \mathrm{Im}\{\big[\widehat{\mathbf{h}}\big]_1\} \nonumber\\
=& 2\left(\left(\mathrm{Re}\{x\}\right)^2+\left(\mathrm{Im}\{x\}\right)^2\right)\,\mathrm{Im}\{\big[\widehat{\mathbf{h}}\big]_1\}.
\end{align}
\end{subequations}
Solving \eqref{eq:real-KP41} in $\mathrm{Re}\{\big[\widehat{\mathbf{h}}\big]_1\}$ and $\mathrm{Im}\{\big[\widehat{\mathbf{h}}\big]_1\}$, and  substituting the resultant into \eqref{eq:h-GLSE-KPb}, yields the desired estimate for $K=1$ as: 
\begin{subequations}
	\begin{eqnarray}\label{eq:h-GLSE-1Pa}
	\mathrm{Re}\{\big[\widehat{\mathbf{h}}\big]_1\}\triangleq\pm\sqrt{\frac{\abs{\left[\overline{\mathbf{H}}_{\mathrm{E}}\right]_{11}}+\mathrm{Re}\{\left[\overline{\mathbf{H}}_{\mathrm{E}}\right]_{11}\}}{2}},
	\end{eqnarray} 
	\begin{eqnarray}\label{eq:h-GLSE-1Pb}
	\mathrm{Im}\{\big[\widehat{\mathbf{h}}\big]_1\}\triangleq\pm
	\frac{\mathrm{Im}\{\left[\overline{\mathbf{H}}_{\mathrm{E}}\right]_{11}\}}{\sqrt{2\left(\abs{\left[\overline{\mathbf{H}}_{\mathrm{E}}\right]_{11}}+\mathrm{Re}\{\left[\overline{\mathbf{H}}_{\mathrm{E}}\right]_{11}\}\right)}},
	\end{eqnarray} 
	\begin{align}\label{eq:h-GLSE-1Pc}
	\left[\begin{array}{cc}
	\mathrm{Re}\{\big[\widehat{\mathbf{h}}\big]_i\} \\
	\mathrm{Im}\{\big[\widehat{\mathbf{h}}\big]_i\} \end{array}\right]\triangleq\frac{\big[\mathbf{Z}_{\mathrm{B}_K}\big]_{i1}\,\mathrm{Re}\{\big[\widehat{\mathbf{h}}\big]_1\}+\big[\mathbf{Z}_{\mathrm{B}_K}\big]_{i2}\,\mathrm{Im}\{\big[\widehat{\mathbf{h}}\big]_1\}}{\abs{\big[\widehat{\mathbf{h}}\big]_i}^2},
	\end{align}
\end{subequations}
$\forall i= 2,3,\cdots,N$. Here $\abs{x}=\sqrt{\left(\mathrm{Re}\{x\}\right)^2+\left(\mathrm{Im}\{x\}\right)^2}$.

\subsubsection{Channel Estimation with full PC,  $K=N$}\label{sec:NP} Here using the fact $\mathbf{E}_N=\mathbf{I}_N$, LSE and LMMSEE for $\mathbf{h}$ are given by:
\begin{equation} 
\widehat{\mathbf{h}}\triangleq  \begin{cases}
\Psi\left\lbrace\Phi\left\lbrace\mathrm{sym}\left\lbrace\widehat{\mathbf{H}}_{\rm L}^*\right\rbrace\right\rbrace\right\rbrace, & \text{LSE,}\\
\Psi\left\lbrace\Phi\left\lbrace\mathrm{sym}\left\lbrace\widehat{\mathbf{H}}_{\rm M}^*\right\rbrace\right\rbrace\right\rbrace, & \text{LMMSEE,}
\end{cases}
\end{equation}  
on using \eqref{eq:h_CE_Final}, along with \eqref{eq:LS-PI2} and  \eqref{eq:hLMMSE} for $K=N$.
Further, with $K=N$, the real and imaginary components of $\widehat{\mathbf{h}}$ can be directly obtained using the maximum eigenvalue  $\lambda_{\mathrm{Z}_1}$ of $\mathbf{Z}$ as:  
\begin{eqnarray}\label{eq:h-GLSE}
\left[\begin{array}{cc}
\mathrm{Re}\{\widehat{\mathbf{h}}\} \\
\mathrm{Im}\{\widehat{\mathbf{h}}\} \end{array}\right]\triangleq\pm\sqrt{\lambda_{\mathrm{Z}_1}}\,\frac{\mathbf{v}_{\mathrm{Z}_1}}{\norm{\mathbf{v}_{\mathrm{Z}_1}}}\in\mathbb{R}^{2N\times1}.
\end{eqnarray}
Here $\mathbf{v}_{\mathrm{Z}_1}\in\mathbb{C}^{N\times1}$ represents the  eigenvector corresponding to the maximum eigenvalue $\lambda_{\mathrm{Z}_1}$ of $\mathbf{Z}$ which is defined below:
\begin{equation} 
\mathbf{Z}\triangleq  \begin{cases}
 \Phi\left\lbrace\mathrm{sym}\left\lbrace\widehat{\mathbf{H}}_{\rm L}^*\right\rbrace\right\rbrace, & \text{LSE,}\\
 \Phi\left\lbrace\mathrm{sym}\left\lbrace\widehat{\mathbf{H}}_{\rm M}^*\right\rbrace\right\rbrace\,\text{ with } K=N, & \text{LMMSEE.}
\end{cases}
\end{equation}

\section{Backscattered SNR Performance Analysis}\label{sec:ana}
\textcolor{black}{In this section we first define the effective average achievable BSC SNR, as denoted by $\overline{\gamma}$, during the ID phase. This metric actually depends on the proposed LSE and LMMSEE based precoder and decoder designs at $\mathcal{R}$.} Thereafter, we also derive the expressions for $\overline{\gamma}$ under the benchmark scenarios of perfect CSI availability and the isotropic transmission from $\mathcal{R}$. Lastly, we conclude the section by presenting a tight analytical approximation of $\overline{\gamma}$, which will be used later for obtaining the joint optimal time allocation (TA) and PC design.

\textcolor{black}{We have adopted the average effective backscattered SNR  $\overline{\gamma}$ as the objective function because the other conventional performance metrics~\cite{inv-BSC-MIMO,Amb-BCS-Ana} like achievable average backscattered throughput and bit error probability during detection are monotonic functions of this $\overline{\gamma}$. So, to maximize the practical efficacy of the proposed CE protocol for BSC, we discourse here the smart multiantenna signal processing to be carried out at $\mathcal{R}$ using the derived closed-form expressions for the jointly-optimal TA and PC design. The performance enhancement achieved in terms of higher BSC range or average backscattered SNR due to this smart selection of TA and PC during CE phase are later numerically characterized in detail in  Section~\ref{sec:comp}.}
\subsection{Average Backscattered SNR received at $\mathcal{R}$ during ID Phase}\label{sec:TxBF}

The maximum array gain is achieved at $\mathcal{R}$ by implementing   MRT to $\mathcal{T}$ in the DL and  
MRC in the UL. 
So, based on the estimate $\widehat{\mathbf{h}}$, the optimal precoder and combiner  are respectively defined
as $\mathbf{g}_{\mathrm{T}}=\frac{\widehat{\mathbf{h}}^*}{\norm{\widehat{\mathbf{h}}^*}}$ and $\mathbf{g}_{\mathrm{R}}=\frac{\widehat{\mathbf{h}}}{\norm{\widehat{\mathbf{h}}}}$. As only $\tau-\tau_c$ is available for ID, the average effective backscattered 
SNR $\overline{\gamma}$ is
\begin{align}\label{eq:gamma1}
\overline{\gamma}\,&\triangleq\;\mathbb{E}\left\lbrace\frac{\left(\tau-\tau_c\right)\,p_t\,\overline{a}^2}{N_0}\left|{\mathbf{g}_{\mathrm{R}}^{\mathrm{H}}\,\mathbf{h}\,\mathbf{h}^{\mathrm{T}}\,\mathbf{g}_{\mathrm{T}}}\right|^2\right\rbrace\nonumber\\
\,&\stackrel{(r1)}{=}\,\frac{\left(\tau-\tau_c\right)p_t\,\overline{a}^2}{N_0}\;\mathbb{E}\left\lbrace\left|\frac{\widehat{\mathbf{h}}^{\mathrm{H}}\,\mathbf{h}}{\norm{\widehat{\mathbf{h}}}}\right|^4\right\rbrace,
\end{align}
where $\overline{a}$ is the average amplitude of the tag's modulation during the ID phase and $(r1)$ is obtained using $\widehat{\mathbf{h}}^{\mathrm{H}}\,\mathbf{h}=\mathbf{h}^{\mathrm{T}}\,\widehat{\mathbf{h}}^*$. 

Now assuming that perfect CSI is available at $\mathcal{R}$, then $\tau_c=0$, i.e., no CE is required, and the optimal precoder and combiner are respectively defined
as $\mathbf{g}_{\mathrm{T}}=\frac{{\mathbf{h}}^*}{\norm{{\mathbf{h}}^*}}$ and $\mathbf{g}_{\mathrm{R}}=\frac{{\mathbf{h}}}{\norm{{\mathbf{h}}}}$. The resulting backscattered SNR is given by:
\begin{align}\label{eq:gammaID} 
\overline{\gamma}_{\rm id}=\frac{\tau\,p_t\,\overline{a}^2}{N_0} \mathbb{E}\left\lbrace\norm{{\mathbf{h}}}^4\right\rbrace \stackrel{(r2)}{=}\frac{\tau\,p_t\,\overline{a}^2}{N_0}N(N+1)\beta^2,
\end{align}
where $(r2)$ is obtained using the fact that $\norm{{\mathbf{h}}}$ follows the  Rayleigh distribution of order $2N$~\cite[eq. 1.12]{simon2007probability}.

On other hand when no CSI is available and no CE is carried out either, then the  effective received backscattered SNR for ID due to the isotropic transmission from $\mathcal{R}$ is given by:
\begin{eqnarray}\label{eq:gammaISO} 
\overline{\gamma}_{\rm is}=\frac{\tau\,p_t\,\overline{a}^2}{N_0}\;\mathbb{E}\left\lbrace\left|\frac{\mathbf{1}_N^{\mathrm{H}}\,\mathbf{h}}{\norm{\mathbf{1}_N}}\right|^4\right\rbrace=\frac{2\,\tau\,p_t\,\overline{a}^2\beta^2}{N_0},
\end{eqnarray} 
where above  is obtained using $\mathbf{g}_{\mathrm{T}}=\mathbf{g}_{\mathrm{R}}=\frac{\mathbf{1}_N}{\norm{\mathbf{1}_N}}$ along with the property that $\sum_{i=1}^{N}[\mathbf{h}]_i$ follows the complex Gaussian distribution with variance $N\beta$ in the following expectation:
\begin{eqnarray}\label{eq:gammaISO0} 
\mathbb{E}\left\lbrace\left|\frac{\mathbf{1}_N^{\mathrm{H}}\,\mathbf{h}}{\norm{\mathbf{1}_N}}\right|^4\right\rbrace= \mathbb{E}\left\lbrace\abs{\frac{\sum_{i=1}^{N}[\mathbf{h}]_i}{\sqrt{N}}}^4\right\rbrace = \frac{2\left(N\beta\right)^2}{N^2}= 2\beta^2.
\end{eqnarray} 

As $\overline{\gamma}$ in \eqref{eq:gamma1} cannot be expressed in closed-form using $\widehat{\mathbf{h}}$ in \eqref{eq:h_CE_Final}, we next present a couple of approximations for the key statistics of $\widehat{\mathbf{h}}$ in Section~\ref{sec:stats} which will be used for obtaining a tight analytical approximation for $\overline{\gamma}$ in Section~\ref{sec:apprG}.

\subsection{Proposed Approximation for Key Statistics of $\widehat{\mathbf{h}}$}\label{sec:stats}
As it is difficult to obtain a closed-form expression for $\overline{\gamma}$, we use a couple of approximations. 
First to obtain the statistics for the conditional $\mathbf{h}\big|\widehat{\mathbf{h}}$ distribution, we use a Gaussian
approximation for the probability density function (PDF) of $\widehat{\mathbf{h}}$~\cite{kay1993fundamentals}. The resulting statistics, the mean and covariance of $\mathbf{h}\big|\widehat{\mathbf{h}}$, under this approximation are respectively given by:
\begin{subequations}\label{eq:Appr1}
	\begin{align}\label{eq:exp0}
	\mathbb{E}\left\lbrace\mathbf{h}\big|\widehat{\mathbf{h}}\right\rbrace\approx&\,\mathbb{E}\left\lbrace\mathbf{h}\right\rbrace+\mathrm{cov}\left(\mathbf{h},\widehat{\mathbf{h}}\right)\left[\mathrm{cov}\left(\widehat{\mathbf{h}},\widehat{\mathbf{h}}\right)\right]^{-1} \widehat{\mathbf{h}},
	\end{align} 
	\begin{align}\label{eq:cov0}
	\mathrm{cov}\left\lbrace\mathbf{h}\big|\widehat{\mathbf{h}}\right\rbrace\approx&\,\mathrm{cov}\left(\mathbf{h},{\mathbf{h}}\right)-\mathrm{cov}\left(\mathbf{h},\widehat{\mathbf{h}}\right)\left[\mathrm{cov}\left(\widehat{\mathbf{h}},\widehat{\mathbf{h}}\right)\right]^{-1}\nonumber\\
	&\,\times \mathrm{cov}\left(\widehat{\mathbf{h}},\mathbf{h}\right).
	\end{align}
\end{subequations}
Now with the LSE of $\mathbf{h}$ as obtained from \eqref{eq:h_CE_Final} being denoted by 
$\widehat{\mathbf{h}}_{\rm L}\triangleq \Psi\left\lbrace\Phi\left\lbrace\mathrm{sym}\left\lbrace\widehat{\mathbf{H}}_{\rm L}^*\,\mathbf{E}_K^{\rm T}\right\rbrace\right\rbrace\right\rbrace$, mean and covariance of $\mathbf{h}\big|\widehat{\mathbf{h}}_{\rm L}$, can be respectively obtained using \eqref{eq:exp0} and \eqref{eq:cov0} as:
\begin{subequations} 
	\begin{align}\label{eq:expL}
	\mathbb{E}\left\lbrace\mathbf{h}\big|\widehat{\mathbf{h}}_{\rm L}\right\rbrace\approx\beta\left[\mathrm{cov}\left(\widehat{\mathbf{h}}_{\rm L},\widehat{\mathbf{h}}_{\rm L}\right)\right]^{-1}\,\widehat{\mathbf{h}}_{\rm L},\quad\text{and} 
	\end{align} 
	\begin{align}\label{eq:covL}
	\mathrm{cov}\left\lbrace\mathbf{h}\big|\widehat{\mathbf{h}}_{\rm L}\right\rbrace\approx\beta\left(\mathbf{I}_N-\beta\left[\mathrm{cov}\left(\widehat{\mathbf{h}}_{\rm L},\widehat{\mathbf{h}}_{\rm L}\right)\right]^{-1}\right).
	\end{align}
\end{subequations}

Likewise with LMMSEE 
$\widehat{\mathbf{h}}_{\rm M} \triangleq \Psi\left\lbrace\Phi\left\lbrace\mathrm{sym}\left\lbrace\widehat{\mathbf{H}}_{\rm M}^*\,\mathbf{E}_K^{\rm T}\right\rbrace\right\rbrace\right\rbrace$, the mean and covariance of $\mathbf{h}\big|\widehat{\mathbf{h}}_{\rm M}$, are respectively  given by:
\begin{subequations} 
	\begin{align}\label{eq:expM}
	\mathbb{E}\left\lbrace\mathbf{h}\big|\widehat{\mathbf{h}}_{\rm M}\right\rbrace\approx\widehat{\mathbf{h}}_{\rm M},\quad\text{and}
	\end{align}
	\begin{align}\label{eq:covM}
	\mathrm{cov}\left\lbrace\mathbf{h}\big|\widehat{\mathbf{h}}_{\rm M}\right\rbrace\approx\beta\, \mathbf{I}_N-\mathrm{cov}\left(\widehat{\mathbf{h}}_{\rm M},\widehat{\mathbf{h}}_{\rm M}\right).
	\end{align}
\end{subequations}

Along with the first one as defined in \eqref{eq:Appr1}, we use the following (second) approximation for the covariance of $\widehat{\mathbf{h}}$:
\begin{align}\label{eq:Appr2}
\mathrm{cov}\left(\widehat{\mathbf{h}},\widehat{\mathbf{h}}\right)=\mathbb{E}\left\lbrace\widehat{\mathbf{h}}\,\widehat{\mathbf{h}}^{\rm H}\right\rbrace \approx\sqrt{\mathbb{E}\left\lbrace\left[\widehat{\mathbf{H}}\right]_{ii}\right\rbrace},
\end{align}
$\forall i=1,2,\ldots,N$, with $K=N$. Here, \eqref{eq:Appr2}  is obtained using the independence and variance of the
zero mean  entries of $\mathbf{h}$ and $\mathbf{W}$ in \eqref{eq:LS-PI}. Using this approximation, the covariance of the LSE and LMMSEE of $\mathbf{h}$ can be respectively approximated as:
\begin{subequations} 
	\begin{equation}\label{eq:covappL}
	\mathrm{cov}\left(\widehat{\mathbf{h}}_{\rm L},\widehat{\mathbf{h}}_{\rm L}\right)=\mathbb{E}\left\lbrace\widehat{\mathbf{h}}_{\rm L}\,\widehat{\mathbf{h}}_{\rm L}^{\rm H}\right\rbrace\approx\sqrt{\beta^2+\frac{N_0}{\mathrm{E}_0}}\;\,\mathbf{I}_N,
	\end{equation}
	\begin{align}\label{eq:covappM}
	\mathrm{cov}\left(\widehat{\mathbf{h}}_{\rm M},\widehat{\mathbf{h}}_{\rm M}\right)=\mathbb{E}\left\lbrace\widehat{\mathbf{h}}_{\rm M}\,\widehat{\mathbf{h}}_{\rm M}^{\rm H}\right\rbrace\approx\,\sqrt{\frac{\beta^4\,{\mathrm{E}_0}}{\beta^2{\mathrm{E}_0}+{N_0}}}\;\,\mathbf{I}_N.
	\end{align} 
\end{subequations}

\subsection{Analytical Approximation for Average Backscattered SNR}\label{sec:apprG}
Using the developments of previous section, here we derive the average BSC SNR $\overline{\gamma}$ during the ID phase using the LSE and LMMSEE for $\mathbf{h}$ as obtained after the CE phase.

\subsubsection{SNR Approximation for LSE}\label{sec:approxL}
Using \eqref{eq:expL}, \eqref{eq:covL}, \eqref{eq:covappL}, we can approximate $\widehat{\mathbf{h}}_{\rm L}$ to follow $\mathbb{C} \mathbb{N}\left(\textbf{0}_{N\times 1},\mathrm{cov}\left(\widehat{\mathbf{h}}_{\rm L},\widehat{\mathbf{h}}_{\rm L}\right)\right)$, which implies that $\norm{\widehat{\mathbf{h}}_{\rm L}}$ can be approximated to follow a Rayleigh distribution of order $2N$. Thus, given $\widehat{\mathbf{h}}_{\rm L}$, the mean and variance for $\varUpsilon_{\rm L}\triangleq\frac{\widehat{\mathbf{h}}_{\rm L}^{\mathrm{H}}\,\mathbf{h}}{\norm{\widehat{\mathbf{h}}_{\rm L}}}$ are  respectively defined by:
\begin{subequations}
	\begin{align}\label{eq:exp2}
	\mu_{{\varUpsilon_{\rm L}}} \triangleq&\,\mathbb{E}\left\lbrace{\varUpsilon_{\rm L}}\big|\widehat{\mathbf{h}}_{\rm L}\right\rbrace=\frac{\widehat{\mathbf{h}}_{\rm L}^{\rm H}\,\mathbb{E}\left\lbrace\mathbf{h}\big|\widehat{\mathbf{h}}_{\rm L}\right\rbrace}{\norm{\widehat{\mathbf{h}}_{\rm L}}} \nonumber\\
	\approx&\,\sqrt{\frac{\beta^2\,\mathrm{E}_0}{\beta^2\,\mathrm{E}_0+N_0}} \norm{\widehat{\mathbf{h}}_{\rm L}}, 
	\end{align}
	\begin{align}\label{eq:var}
	\sigma^2_{{\varUpsilon_{\rm L}}}\triangleq\mathrm{var}\left\lbrace{\varUpsilon_{\rm L}}\big|\widehat{\mathbf{h}}_{\rm L}\right\rbrace\approx\beta\left(1-\sqrt{\frac{\beta^2\,\mathrm{E}_0}{\beta^2\,\mathrm{E}_0+N_0}}\right).
	\end{align}
\end{subequations}
So, with a Gaussian approximation for the PDF of $\widehat{\mathbf{h}}_{\rm L}$, ${\varUpsilon_{\rm L}}\big|\widehat{\mathbf{h}}_{\rm L}\sim\mathbb{CN}\left(\mu_{{\varUpsilon_{\rm L}}},\sigma^2_{{\varUpsilon_{\rm L}}}\right)$, and hence $\abs{{\varUpsilon_{\rm L}}}\big|\widehat{\mathbf{h}}_{\rm L}$ follows the Rician distribution. Thus, on using the fourth moment of $\abs{{\varUpsilon_{\rm L}}}\big|\widehat{\mathbf{h}}_{\rm L}$ in \eqref{eq:gamma1}, we obtain the desired approximation $\overline{\gamma}_{\rm La}$ for the average BSC SNR $\overline{\gamma}_{\rm L}$ for ID using the LSE $\widehat{\mathbf{h}}_{\rm L}$ as:
\begin{align}\label{eq:GaL}
&\overline{\gamma}_{\rm L}\triangleq\frac{\left(\tau-\tau_c\right)p_t\,\overline{a}^2}{N_0}\;\mathbb{E}\left\lbrace\left|\frac{\widehat{\mathbf{h}}_{\rm L}^{\mathrm{H}}\,\mathbf{h}}{\norm{\widehat{\mathbf{h}}_{\rm L}}}\right|^4\right\rbrace\approx\nonumber\\
&\overline{\gamma}_{\rm La}\stackrel{(r3)}{\triangleq}\frac{\left(\tau-\tau_c\right)p_t}{N_0\left(\overline{a}\right)^{-2}}\,\mathbb{E}_{\widehat{\mathbf{h}}_{\rm L}}\!\!\left\lbrace\left(\mu_{{\varUpsilon_{\rm L}}}\right)^4+4 \left(\mu_{{\varUpsilon_{\rm L}}}\right)^2 \sigma^2_{{\varUpsilon_{\rm L}}}+2 \left(\sigma^2_{{\varUpsilon_{\rm L}}}\right)^2\right\rbrace\nonumber\\
&\stackrel{(r4)}{=}\frac{\left(\tau-\tau_c\right)p_t}{N_0\left(\overline{a}\beta\right)^{-2}}\left(\frac{N^2-3 N+2}{1+\frac{N_0\,K}{\beta^2a_0^2p_t\tau_c}} +\frac{4 (N-1)}{\sqrt{1+\frac{N_0\,K}{\beta^2a_0^2p_t\tau_c}}}+2\right).
\end{align}
Here $(r3)$ uses $4$th moment of Rician variable of order $2N$~\cite[eq. 2.23]{simon2007probability} in  $\mathbb{E}\left\lbrace\left|{\varUpsilon_{\rm L}}\right|^4\right\rbrace=\mathbb{E}_{\widehat{\mathbf{h}}_{\rm L}}\left\lbrace \mathbb{E}_{\mathbf{h}\big|\widehat{\mathbf{h}}_{\rm L}}\left\lbrace\left|{\varUpsilon_{\rm L}}\right|^4\right\rbrace\right\rbrace$. Whereas,  $(r4)$ is obtained using  $\mathbb{E}\left\lbrace\norm{\widehat{\mathbf{h}}_{\rm L}}^2\right\rbrace\approx N\sqrt{\beta^2+\frac{N_0\,K}{a_0^2\,p_t\,\tau_c}}$ and $\mathbb{E}\left\lbrace\norm{\widehat{\mathbf{h}}_{\rm L}}^4\right\rbrace\approx N(N+1)\left(\beta^2+\frac{N_0\,K}{a_0^2\,p_t\,\tau_c}\right)$.

\subsubsection{SNR Approximation for LMMSEE}\label{sec:approxM}
Using \eqref{eq:expM}, \eqref{eq:covM}, and \eqref{eq:covappM}, the mean $\mu_{\varUpsilon_{\rm M}}$ and variance $\sigma^2_{\varUpsilon_{\rm M}}$ for $\varUpsilon_{\rm M}\triangleq\frac{\widehat{\mathbf{h}}_{\rm M}^{\mathrm{H}}\,\mathbf{h}}{\norm{\widehat{\mathbf{h}}_{\rm M}}}$ for a given LMMSEE $\widehat{\mathbf{h}}_{\rm M}$ are respectively approximated as:
\begin{subequations}
	\begin{eqnarray}\label{eq:exp2M}
	\mu_{\varUpsilon_{\rm M}}\triangleq\mathbb{E}\left\lbrace\varUpsilon_{\rm M}\big|\widehat{\mathbf{h}}_{\rm M}\right\rbrace=\frac{\widehat{\mathbf{h}}_{\rm M}^{\rm H}\,\mathbb{E}\left\lbrace\mathbf{h}\big|\widehat{\mathbf{h}}_{\rm M}\right\rbrace}{\norm{\widehat{\mathbf{h}}_{\rm M}}} \approx  \norm{\widehat{\mathbf{h}}_{\rm M}}\;\text{ and}
	\end{eqnarray}
	\begin{align}\label{eq:varM}
	\sigma^2_{\varUpsilon_{\rm M}}\triangleq\mathrm{var}\left\lbrace\varUpsilon_{\rm M}\big|\widehat{\mathbf{h}}_{\rm M}\right\rbrace\approx\beta -\sqrt{\frac{\beta^4\,{\mathrm{E}_0}}{\beta^2\,{\mathrm{E}_0}+{N_0}}}.
	\end{align}
\end{subequations}

Hence, with a Gaussian approximation for the PDF of $\widehat{\mathbf{h}}_{\rm M}$, $\varUpsilon_{\rm M}\big|\widehat{\mathbf{h}}_{\rm M}\sim\mathbb{CN}\left(\mu_{\varUpsilon_{\rm M}},\sigma^2_{\varUpsilon_{\rm M}}\right)$, we notice that $\abs{\varUpsilon_{\rm M}}\big|\widehat{\mathbf{h}}_{\rm M}$ follows the Rician distribution. Thus, on using the fourth moment of $\abs{\varUpsilon_{\rm M}}\big|\widehat{\mathbf{h}}_{\rm M}$ in \eqref{eq:gamma1}, the approximation $\overline{\gamma}_{\rm Ma}$ for BSC SNR $\overline{\gamma}_{\rm M}=\frac{\left(\tau-\tau_c\right)\,p_t\,\overline{a}^2}{N_0}\;\mathbb{E}\left\lbrace\left|\frac{\widehat{\mathbf{h}}_{\rm M}^{\mathrm{H}}\,\mathbf{h}}{\norm{\widehat{\mathbf{h}}_{\rm M}}}\right|^4\right\rbrace$ using LMMSEE $\widehat{\mathbf{h}}_{\rm M}$ is given by:
\begin{align}\label{eq:GaM}
\overline{\gamma}_{\rm Ma} \stackrel{(r5)}{\triangleq}\frac{\left(\tau-\tau_c\right)\,p_t\,\overline{a}^2\beta ^2}{N_0}&\left(\frac{\beta}{\sigma^2_{\widehat{\mathrm{h}}}}\left(N-1\right)\times\right.\nonumber\\ &\left.\left(\frac{\beta}{\sigma^2_{\widehat{\mathrm{h}}}}\left(N-2\right)+4\right)+2\right),
\end{align}
where $\sigma^2_{\widehat{\mathrm{h}}}\triangleq\sqrt{\beta^2+\frac{K\,N_0}{a_0^2\,\mathrm{E}_\mathrm{c}}}$ and $(r5)$ is obtained using the following two key results along with \eqref{eq:exp2M} and \eqref{eq:varM}:
\begin{subequations}
	\begin{align}\label{eq:covappM2}
	\mathbb{E}\left\lbrace\widehat{\mathbf{h}}_{\rm M}^{\rm H}\,\widehat{\mathbf{h}}_{\rm M}\right\rbrace=\mathbb{E}\left\lbrace\norm{\widehat{\mathbf{h}}_{\rm M}^{\rm H}}^2\right\rbrace\approx N \sqrt{\frac{\beta^4\,{\mathrm{E}_0}}{\beta^2\,{\mathrm{E}_0}+{N_0}}}, 
	\end{align}
	\begin{align}\label{eq:covappM4}
	\mathbb{E}\left\lbrace\norm{\widehat{\mathbf{h}}_{\rm M}^{\rm H}}^4\right\rbrace\approx N\left(N+1\right)\left(\frac{\beta^4\,{\mathrm{E}_0}}{\beta^2\,{a_0^2\,\mathrm{E}_\mathrm{c}}+{N_0}}\right).
	\end{align}
\end{subequations}
Since from \eqref{eq:GaL} and \eqref{eq:GaM} we notice that $\overline{\gamma}_{\rm La}=\overline{\gamma}_{\rm Ma}$, we denote the approximated effective BSC SNR by $\overline{\gamma}_{\rm a}\triangleq\overline{\gamma}_{\rm La}=\overline{\gamma}_{\rm Ma}$.


\section{Joint Resource Optimization at Reader}\label{sec:opt}
This section is dedicated towards the joint optimization study for finding the most efficient utilization of the energy available at $\mathcal{R}$ for CE and ID along with the smart selection of the orthogonal PC $K$ for obtaining the LSE or LMMSEE of $\mathbf{h}$. We start with individually optimizing energy and PC, before proceeding with the joint optimization in the last part.

\subsection{Optimal Energy Allocation at Reader for CE and ID}
First we focus on optimally distributing the energy at $\mathcal{R}$ between the CE and ID phases. Assuming a given transmit power, fixed at the maximum level $p_t$ and $\tau_c=KL$ in seconds, we find this energy allocation by optimizing the length $L$ of the  pilots to decide on the
TA $\tau_c$ for the CE phase and $\tau-\tau_c$ for the ID phase. Next after proving the quasiconcavity of the optimization metric  $\overline{\gamma}$ $\left(\text{or } \overline{\gamma}_{\rm a}\right)$ in TA $\tau_c$ for CE phase to enable efficient ID using the LSE $\widehat{\mathbf{h}}_{\rm L}$ or LMMSEE $\widehat{\mathbf{h}}_{\rm M}$, we present a tight analytical approximation for global optimal  $\tau_c$.

\textcolor{black}{Before proceeding with the optimal TA scheme, we would like to highlight that the objective function $\overline{\gamma}_{\rm a}$ (cf. \eqref{eq:GaL}) to be maximized being non-decreasing in $p_t$, i.e., $\frac{\partial\overline{\gamma}_{\rm a}}{\partial p_t}\ge0$, is the reason behind selection of optimal power allocation strategy of equally distributing entire  power budget $p_t$ over the transmitting antennas at $\mathcal{R}$.}

\subsubsection{Quasiconcavity of SNR $\overline{\gamma}$ in $\tau_c$}\label{sec:quasi}
As the LSE $\widehat{\mathbf{h}}_{\rm L}$ or LMMSEE $\widehat{\mathbf{h}}_{\rm M}$ cannot be obtained in closed-form due to the involvement of eigenvalue decomposition defined in \eqref{eq:h_CE_Final},  we   analyze the properties of $\overline{\gamma}$ as a function of $\tau_c$ under CE errors in an alternate way. 
With $\mathrm{E}_\mathrm{c}\triangleq p_t\,\tau_c$, from \eqref{eq:LS-PI} we notice that the role of $\tau_c$ in the CE phase is to bring $\widehat{\mathbf{H}}$ as close as  possible to $\mathbf{H}_K$ (i.e., minimize $\Theta\left\lbrace\widehat{\mathbf{H}}\right\rbrace$ in $\mathcal{OP}_K$), while leaving sufficient time $\left(\tau-\tau_c\right)$ for ID. 
\textcolor{black}{So, there exists a tradeoff between the CE quality improvement by having larger CE time $\tau_c$ and spectral efficiency enhancement by leaving a larger fraction of the coherence time
dedicated for carrying out ID.} With $a_0^2,p_t\ge0,$ the distance between $\widehat{\mathbf{H}}$ and  $\mathbf{H}_K$, \Big(for example, $\mathbb{E}\left\lbrace\norm{\widehat{\mathbf{H}}_{\rm L}-\mathbf{H}_K}^2\right\rbrace=\frac{N_0}{a_0^2\,p_t\tau_c}$\Big), is monotonically decreasing in $\tau_c$ and attains its
minimum (i.e., zero) only when either $\mathrm{E}_\mathrm{c}=p_t\tau_c\to\infty$ or $N_0\to0$. Moreover, the rate of this decrease (i.e.,  improvement in CE quality) is diminishing in $\tau_c$.   Since, $\overline{\gamma}$, regardless of the underlying conditional distribution of $\mathbf{h}$ for a given  $\widehat{\mathbf{h}}$, is a monotonically decreasing function of this distance or error in CE, 
$\overline{\gamma}$ is monotonically non-decreasing in $\tau_c$, with this rate of increase with $\tau_c$ being non-increasing. Combining this observation with the result in the following lemma   proves the quasiconcavity~\cite{Baz} of $\overline{\gamma}$ in $\tau_c$.
\begin{lemma}\label{lem:1}
	\textit{For a non-decreasing positive function $\mathcal{B}\left(x\right)$ whose rate of increase is non-increasing, the product   $\mathcal{A}\left(x\right)\triangleq\left(1-x\right)\mathcal{B}\left(x\right)$ is quasiconcave in $x$, $\forall\, 0\le x\le 1$.}
\end{lemma}
\begin{IEEEproof}
	If $\exists$ $x^*\triangleq\left\lbrace x \Big| \frac{\partial\mathcal{A}\left(x\right)}{\partial x}= \left(1-x\right)\frac{\partial\mathcal{B}\left(x\right)}{\partial x}-\mathcal{B}\left(x\right)=0\right\rbrace$, then it can be observed that $\mathcal{B}\left(x\right)>\left(1-x\right)\frac{\partial\mathcal{B}\left(x\right)}{\partial x},\forall x>x^*,$ using the properties of $\mathcal{B}$. This along with $\mathcal{A}\left(x\right)=0$ for $x=1$, completes the proof for quasiconcavity of $\overline{\gamma}$ in $\tau_c$.
\end{IEEEproof}

\subsubsection{Analytical Approximation for Global Optimal $\tau_c$}\label{sec:approx-GOP}
Firstly, its worth noting that since $\frac{\partial^2\overline{\gamma}_{\rm a}}{\partial \tau_c^2}\le0$, it implies \textit{concavity} of $\overline{\gamma}_{\rm a}$ in $\tau_c$. This corroborates the general unimodality claim made in Lemma~\ref{lem:1}, and exploiting these results, a tight approximation $\tau_{c{\rm{a}}}$ for the global optimal $\tau_c$ can be obtained using any root finding technique or the bisection method for solving $\frac{\partial\overline{\gamma}_{\rm a}}{\partial \tau_c}=0$ in $\tau_c$, which is a \textit{quintic} function (a polynomial of degree five). So, $\tau_{c{\rm{a}}}\triangleq\left\lbrace \tau_c \,\Big|\,\frac{\partial\overline{\gamma}_{\rm a}}{\partial \tau_c}=0\right\rbrace$. Here we would like to remind that for univariate functions, unimodality and quasiconcavity are equivalent~\cite{Baz}, and concave functions are quasiconcave also.

Hence, from $\tau_{c{\rm{a}}}$, the total energy budget $\mathrm{E}_{\mathrm{tot}}\triangleq p_t\,\tau$ at $\mathcal{R}$ can be optimally distributed between the CE and ID phases as $p_t\,\tau_{c{\rm{a}}}$ and $p_t\,\left(\tau-\tau_{c{\rm{a}}}\right)$, respectively, to maximize $\overline{\gamma}$ in~\eqref{eq:gamma1}.

\subsection{Optimal Orthogonal Pilots Count $K$ during CE Phase}\label{sec:optK}
To find optimal PC, as denoted by $K_{opt}$, for the orthogonal pilots to be used during CE that can yield the maximum $\overline{\gamma}_{\rm a}$ for a given $\tau_c$,  we first present a key convexity property as obtained after relaxing integer constraint on $K\in\{1,2,\ldots,N\}$.
\begin{lemma}\label{lem:cvx}
	The proposed tight approximation for the average backscattered SNR $\overline{\gamma}_{\rm a}$ is convex in integer-relaxed PC $K\in\mathbb{R}$.
\end{lemma}
\begin{IEEEproof}
	Approximated SNR $\overline{\gamma}_{\rm a}$ can also be represented as:
	\begin{equation}
	\overline{\gamma}_{\rm a}\triangleq\left(\mathrm{f_o}\circ\mathrm{f_i}\right)\left(K\right)=\mathrm{f_o}\left(\mathrm{f_i}\left(K\right)\right),
	\end{equation}
	where   $\mathrm{f_o}\left(x\right)=\frac{\left(\tau-\tau_c\right)\,p_t\,\overline{a}^2\beta ^2}{N_0} \Big(\frac{\beta}{x}\left(N-1\right) \left(\frac{\beta}{x}\left(M-2\right)+4\right)+2\Big)$ and $\mathrm{f_i}\left(K\right)=\sqrt{\beta^2+\frac{K\,N_0}{a_0^2\,\mathrm{E}_\mathrm{c}}}$. Now here we notice that $\frac{\partial^2\mathrm{f_i}\left(K\right)}{\partial K^2}\le0$  and $\frac{\partial^2\mathrm{f_o}\left(x\right)}{\partial x^2}\ge0$ with $\frac{\partial\mathrm{f_o}\left(x\right)}{\partial x}\le0$, respectively implies the concavity of $\mathrm{f_i}$ in continuous $K$ and non-increasing convexity of $\mathrm{f_o}$ in $x$. So, as the non-increasing convex transformation of a concave function is convex~\cite[eq. (3.10)]{boyd}, the convexity of $\overline{\gamma}_{\rm a}$ in integer-relaxed  PC $K$ is hence proved.
\end{IEEEproof}

As we intend to maximize   $\overline{\gamma}_{\rm a}$, which is convex in $K$ under the integer relaxation, the optimal $K$ has to be defined by either of the two corner points, i.e., $K_{\rm opt}=1$ or $K_{\rm opt}=N$. The latter holds because the conner points yield the maxima for a convex function. This decision on which corner point to be selected is based on a SNR threshold $\overline{\gamma}_{\rm th}$ as defined below:
\begin{equation}\label{eq:Kopt}
K_{\rm opt}\triangleq\begin{cases}
1, & \text{$\overline{\gamma}_{\rm E}\triangleq\frac{\beta^2\,a_0^2\,\mathrm{E}_\mathrm{c}}{N_0}\le\overline{\gamma}_{\rm th}\triangleq\frac{\left(N-1\right)^2}{8\left(N+1\right)},$}\\
N, & \text{otherwise},
\end{cases}
\end{equation}   
which has been obtained after finding out whether the underlying approximate CE error $\left(\beta-\sqrt{\beta^2+\frac{K\,N_0}{a_0^2\,\mathrm{E}_\mathrm{c}}}\right)^2$ is lower with $K=1$ or for $K=N$. 

\color{black}\begin{proposition}
	Using \eqref{eq:Kopt}, we can make two observations:
	
	\noindent (a) with massive antenna array (i.e., $ N\gg1$) at $\mathcal{R}$,   $K_{\rm opt}=1$,
	
	\noindent (b) for high SNR scenarios having $\overline{\gamma}_{\rm E}\gg1$, $K_{opt}=N$. 
\end{proposition}
\begin{IEEEproof}
(a) For the massive antenna array at $\mathcal{R}$, the definition of $\overline{\gamma}_{\rm th}$ implies that $\overline{\gamma}_{\rm th}\gg1,\forall\, N\gg1$. Therefore, $\overline{\gamma}_{\rm th}>\overline{\gamma}_{\rm E}$, and hence, optimal $K$ will be always $1$,
This happens because with increasing $N$ at $\mathcal{R}$, the transmit power $\frac{p_t}{N}$ over each antenna keeps on decreasing.

(b) On other hand for high SNR scenarios, implying $\overline{\gamma}_{\rm E}\gg1$, $K_{opt}=N$ because $\overline{\gamma}_{\rm E}$ here is generally higher than $\overline{\gamma}_{\rm th}$.
\end{IEEEproof} 

Below we discuss the physical interpretations behind \eqref{eq:Kopt}.
\begin{remark}\label{rem:ins-optK}
	The intuition for convexity of $\overline{\gamma}_{\rm a}$ in $K$, that eventually resulted in its optimal value defined in  \eqref{eq:Kopt}, is the underlying tradeoff between having larger lower-quality samples  available for CE versus to have fewer better-quality samples. Hence, when the channel conditions are favorable, i.e.,  $\overline{\gamma}_{\rm E}>\overline{\gamma}_{\rm th}$, having $N^2$ lower-quality samples at $\mathcal{R}$ during CE due to lower transmit power $\frac{p_t}{N}$ over each antenna for $K=N$ setting is preferred over having $N$ better-quality backscattered samples with entire transmit power budget $p_t$ allocated to the only antenna transmitting for $K=1$ case.
\end{remark}

\begin{remark}\label{rem:Phy-Int}
	Another key insight for this nontrivial property of the optimal PC $K_{\rm opt}$ stems from the definition for  $\widehat{\mathbf{h}}$ given in \eqref{eq:hhatK}. Since, the accuracy of CE for the last $N-K$ entries $\mathbf{h}_{\bar{K}}$ (cf. \eqref{eq:h-GLSE-KPb}) depends on the quality of estimate for the first $K$ entries  $\mathbf{h}_K$ (cf. \eqref{eq:h-GLSE-KPa}), $K_{\rm opt}=N$ when $\widehat{\mathbf{h}}$ in \eqref{eq:h-GLSE} is accurate enough based on the underlying average SNR $\overline{\gamma}_{\rm E}$ value during CE being greater than the threshold $\overline{\gamma}_{\rm th}$. Otherwise, its better to choose $K=1$ over $K>1$ because this inaccuracy in estimating $\mathbf{h}_K$ also adversely affects the quality of the remaining $N-K$ estimates as denoted by $\mathbf{h}_{\bar{K}}$.
\end{remark}\color{black}

\subsection{Joint Energy Allocation and PC for Maximizing $\overline{\gamma}_{\rm a}$}\label{sec:joint}
With transmit power set to the maximum permissible value $p_t$, the problem of joint energy allocation $p_t\,\tau_c$ and PC $K$ for CE to maximize $\overline{\gamma}_{\rm a}$ can be mathematically formulated as:
\begin{equation*}\label{eqOPT}
\begin{split}
\mathcal{J}:\;&\underset{\tau_c,K}{\rm maximize} \;\; \overline{\gamma}_{\rm a}=\frac{\frac{\left(\tau-\tau_c\right)\,p_t\,\overline{a}^2}{N_0} \bigg(\frac{\beta\left(N-2\right)}{\sqrt{\beta^2+\frac{K\,N_0}{a_0^2\,p_t\,\tau_c}}}+4\bigg)}{\frac{1}{\beta^3\left(N-1\right)}\sqrt{\beta^2+\frac{K\,N_0}{a_0^2\,p_t\,\tau_c}}} +2\beta^2,\\
&\textrm{subject to}
~({\rm C2}):0\leq\tau_c\leq\tau,
~({\rm C3}):K\in\{1,2,\cdots,N\}.
\end{split}
\end{equation*}
As $\mathcal{J}$ is a combinatorial nonconvex problem, we present an alternate methodology to obtain its joint optimal solution as denoted by $\left(\tau_{c,\rm jo},K_{\rm jo}\right)$. In this regard, as  from \eqref{eq:Kopt} the optimal PC satisfies $K_{\rm jo}=1$ or $K_{\rm jo}=N$, below we first define the underlying optimal TA $\tau_{c{\rm{a}_1}}$ for $K=1$ and $\tau_{c{\rm{a}_N}}$ for $K=N$:
\begin{equation}\label{eq:topti}
	\tau_{c{\mathrm{a}_i}}\triangleq\left\lbrace \tau_c \,\Big|\,\left(\frac{\partial\overline{\gamma}_{\rm a}}{\partial \tau_c}=0\right)\wedge\left(K=i\right)\right\rbrace\,\forall i=\{1,N\}.
\end{equation} 
Here, we recall that $\tau_{c{\mathrm{a}_i}}<\tau_{c{\rm{a}_N}}$, which has also been validated later via numerical results plotted in Figs.~\ref{fig:tc_var_LMMSE} and~\ref{fig:OTA_OPC_Nvar}, because more entries of $\mathbf{H}_K$ needs to be estimated for $K=N$ (i.e., $N^2$ entries from $N\times N$ received matrix) than for $K=1$ ($N$ entries from $N\times 1$ received vector). Using  this information in \eqref{eq:Kopt}, the optimal $K$ for $\mathcal{J}$ can be defined as:
\begin{equation}\label{eq:Koptj}
K_{\rm jo}\triangleq\begin{cases}
1, & \text{$\overline{\gamma}_{\rm E_1}\triangleq\frac{\beta^2\,a_0^2\,p_t\,\tau_{c{\rm{a}_1}}}{N_0}\le\frac{\left(N-1\right)^2}{8\left(N+1\right)}$},\\
N, & \text{otherwise}.
\end{cases}
\end{equation} 
Substituting $K_{\rm jo}$ in \eqref{eq:topti}, the desired optimal TA $\tau_{c,\rm jo}$ in $\mathcal{J}$ is:
\begin{equation}\label{eq:toptj}
\tau_{c,\rm jo}\triangleq\left\lbrace \tau_c \,\Big|\,\left(\frac{\partial\overline{\gamma}_{\rm a}}{\partial \tau_c}=0\right)\wedge\left(K=K_{\rm jo}\right)\right\rbrace.
\end{equation} 
Hence, the analytical expressions in \eqref{eq:toptj} and \eqref{eq:Koptj} yield the desired joint sub-optimal TA and PC solution for the nonconvex combinatorial problem $\mathcal{J}$. \textcolor{black}{These closed-form expressions not only provide key analytical design insights, but also incur very low computational cost at $\mathcal{R}$.} Extensive simulation results have been provided in next section to validate the quality of this proposed joint solution along with the quantification of the achievable gains on using it over the fixed benchmark schemes. 

\begin{figure}[!t]
	\centering 
	\includegraphics[width=3.48in]{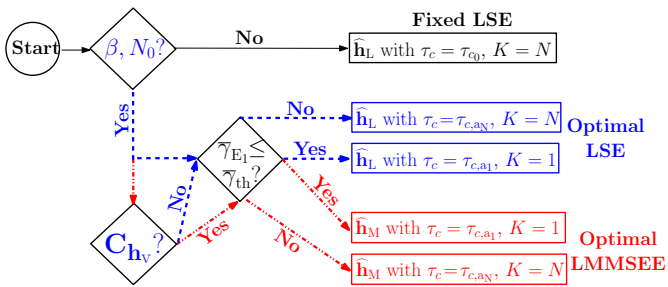}
	\caption{\small The decision tree summarizing the joint optimal PC, energy allocation, and CE technique selection, with fixed TA denoted by $\tau_{c_0}$.}
	\label{fig:decision} 
\end{figure} 
\begin{remark}
	The decision making for obtaining joint optimal energy allocation $p_t\,\tau_c$ and PC $K$ for CE along with selection of LSE $\widehat{\mathbf{h}}_{\rm L}$ and LMMSEE $\widehat{\mathbf{h}}_{\rm M}$ has been summarized in Fig.~\ref{fig:decision}. So, we notice that based on the availability of information on the key parameters $\beta,N_0,\mathbf{C}_{\mathbf{h}_{\rm v}}, N,$ and the relative value of average SNR $\overline{\gamma}_{\rm E_1}$ during CE phase  with $K=1$, the optimal CE technique and resource allocation can be decided to yield a tight approximate for the global maximum value of $\overline{\gamma}$.
\end{remark}

\section{Numerical Results}\label{sec:res}
Here we conduct a  detailed numerical investigation to validate the proposed estimates for the backscattered channel, average SNR performance analysis, and the joint optimization results. Unless explicitly stated, we have used $N=20, K=N, \tau_c=\tau_{c_0}=0.1$ms with $L=5\;\mu$s~\cite{RFID-bookch}, $\tau=1$ ms, $p_t=30$ dBm, $a_0=0.78$~\cite{BSC-Cascaded}, $\overline{a}=0.3162$~\cite{Bistatic-BCS}, and $\beta=\frac{\left(3\times 10^8\right)^2}{\left(4\pi f\right)^2 d^{\varrho}},$ where $f=915$ MHz is the carrier frequency, and 
$d=100$ m with $\varrho=2.5$ as path loss exponent. The AWGN variance is set to $N_0=k_B\,T\,10^{F/10}\approx10^{-20}$ J, where $k_B=1.38\times10^{-23}$ J/K, $T=300$ K, and the noise figure is $F=7$dB. All the simulation results plotted here have been obtained numerically after averaging over $10^5$ independent channel realizations.

\subsection{Validation of the Proposed CE and SNR Analysis}\label{sec:valid}
Here first we validate the quality of the proposed LSE and LMMSEE for $\mathbf{h}$ using both $K=N$ and $K=1$ orthogonal pilots transmission from $\mathcal{R}$ during the CE phase. After that we focus on verifying the tightness of the derived closed-form approximation $\overline{\gamma}_{\rm a}$ for the average BSC SNR $\overline{\gamma}$ during ID phase which has been used for obtaining joint optimal TA and PC.
\begin{figure}[!t]
	\centering 
	\includegraphics[width=3.55in]{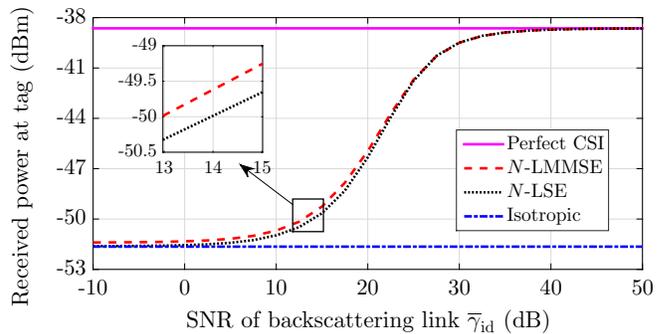}
	\caption{\small Validating the quality  of LSE $\widehat{\mathbf{h}}_{\rm L}$ and LMMSEE $\widehat{\mathbf{h}}_{\rm M}$ with PC $K=N$ in terms of average received power at $\mathcal{T}$ for different SNRs available for ID with perfect CSI at $\mathcal{R}$. Performances for perfect CSI-based and isotropic transmissions are also plotted as benchmarks.}
	\label{fig:LSE-N} 
\end{figure} 
\subsubsection{Validating the proposed CE quality}\label{sec:valid-CE} Considering $K=N$ orthogonal pilots for CE, via Fig.~\ref{fig:LSE-N} we verify the performance of proposed LSE $\widehat{\mathbf{h}}_{\rm L}$ and LMMSEE $\widehat{\mathbf{h}}_{\rm M}$ (cf. \eqref{eq:h_CE_Final}) against increasing average backscattered SNR $\overline{\gamma}_{\rm id}$ (cf. \eqref{eq:gammaISO}) during the ideal scenario of having perfect CSI availability at $\mathcal{R}$. With the average received RF power $p_r\triangleq p_t \,\mathbb{E}\left\lbrace\left|\frac{\widehat{\mathbf{h}}^{\mathrm{H}}\,\mathbf{h}}{\norm{\widehat{\mathbf{h}}}}\right|^2\right\rbrace$ with $K=N$ in $\widehat{\mathbf{h}}$ being the performance validation metric for estimating the goodness of  $\widehat{\mathbf{h}}_{\rm L}$ and $\widehat{\mathbf{h}}_{\rm M}$, we have also plotted the perfect CSI (no CE error) and isotropic (no CSI required) transmission  cases to respectively give   upper and lower bounds on $p_r$. The average received powers for the perfect-CSI and isotropic transmission cases are respectively given by $p_t \,\mathbb{E}\left\lbrace\left|\frac{\mathbf{h}^{\mathrm{H}}\,\mathbf{h}}{\norm{\mathbf{h}}}\right|^2\right\rbrace=N\,p_t\,\beta$ and $p_t \,\mathbb{E}\left\lbrace\left|\frac{\mathbf{1}_N^{\mathrm{H}}\,\mathbf{h}}{\norm{\mathbf{1}_N}}\right|^2\right\rbrace=p_t\,\beta$, where $\mathbf{1}_N$ is an all-one $N\times1$ vector. As observed from Fig.~\ref{fig:LSE-N}, the quality of both proposed LSE and LMMSEE improve with increasing SNR $\overline{\gamma}_{\rm id}$ because the underlying CE errors reduce, and for $\overline{\gamma}_{\rm id}>35$dB, the corresponding $p_r$ approaches $N\,p_t\,\beta$, i.e., the performance   achieved with perfect CSI availability. Further, $\widehat{\mathbf{h}}_{\rm M}$ yields a better CE as compared to $\widehat{\mathbf{h}}_{\rm L}$ with an average performance gap of $-93$dB  between them for $\overline{\gamma}_{\rm id}$ ranging from $-10$dB to $60$dB. However, for $\overline{\gamma}_{\rm id}>25$dB, LSE and LMMSEE yield a very similar performance in $p_r$ at $\mathcal{T}$. 

\begin{figure}[!t]
	\centering 
	\includegraphics[width=3.6in]{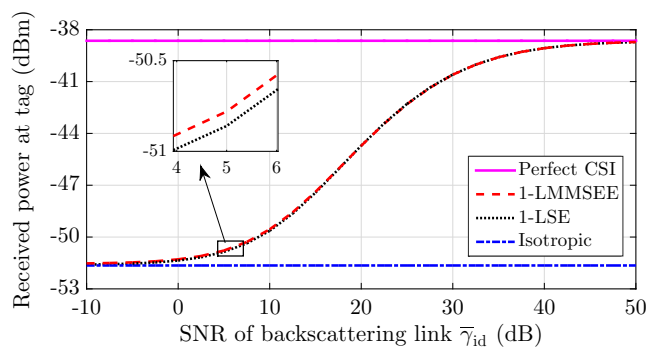}
	\caption{\small Verifying the quality of the LSE and LMMSEE for $\mathbf{h}$ under single pilot $(K=1)$ transmission from $\mathcal{R}$ with different SNR values.}
	\label{fig:LSE-1} 
\end{figure}
Next we investigate the impact of considering a single pilot $K=1$ transmission from $\mathcal{R}$ during the CE. From Fig.~\ref{fig:LSE-1}, we notice a similar trend in the quality of $\widehat{\mathbf{h}}_{\rm M}$ and $\widehat{\mathbf{h}}_{\rm L}$ being getting enhanced with increasing $\overline{\gamma}_{\rm id}$. However, the performance gap between LMMSEE and LSE for $\mathbf{h}$ in terms of $p_r$ is reduced to about $-100$dB for $K=1$. Also, for $K=1$ the $p_r$ for the two estimates approaches to $N p_t \beta$ for relatively higher SNRs values, i.e., $\overline{\gamma}_{\rm id}>45$dB. But in contrast, the  average receiver power $p_r$ performance at $\mathcal{T}$ in the low SNR regime, i.e., $-10\text{dB}\le\overline{\gamma}_{\rm id}\le10$dB is better for $K=1$ as sown in Fig.~\ref{fig:LSE-1} in comparison to that with $K=N$ in Fig.~\ref{fig:LSE-N}. More insights on these results  are presented later in Section~\ref{sec:res-OptK}. 

\begin{figure}[!t]
	\centering\color{black} 
	\includegraphics[width=3.48in]{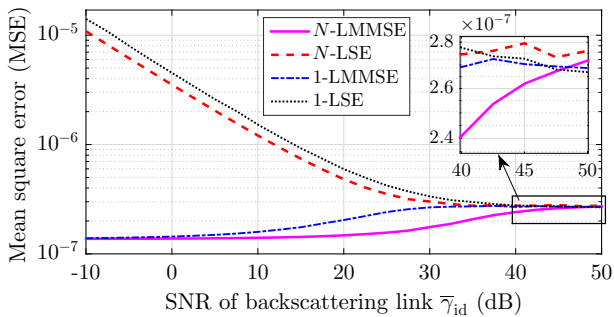}
	\caption{\small  MSE between the  actual channel vector $\mathbf{h}$ and the proposed (LS and LMMSE based) estimates  $\widehat{\mathbf{h}}$ for different SNR $\overline{\gamma}_{\rm id}$ values.}
	\label{fig:MSE}\color{black} 
\end{figure}
\textcolor{black}{We conclude the validation of proposed LSE and LMMSEE quality by plotting the conventional mean square error (MSE)~\cite{massive-MIMO} between the actual channel $\mathbf{h}$ and its estimate $\widehat{\mathbf{h}}$ in Fig.~\ref{fig:MSE}. Noting that the MSE for both our LS and LMMSE based estimates is $<10^{-6}$ in most of the SNR regime, this result verifies the accuracy of our proposed CE paradigms for BSC as discoursed in Sections~\ref{sec:form} and~\ref{sec:est}.  This result is also presented to support the preference of received power $p_r$ as validation metric over the MSE. Actually, since our proposed estimates, as defined in \eqref{eq:h_CE_Final}, are unable to resolve the underlaying phase ambiguity (cf. \eqref{eq:h-GLSE-KPa}), it becomes critical to consider a performance validation metric that can also incorporate the resulting phasor mismatch between $\mathbf{h}$ and $\widehat{\mathbf{h}}$, other than their magnitude difference. As $p_r= p_t \,\mathbb{E}\left\lbrace\left|\frac{\widehat{\mathbf{h}}^{\mathrm{H}}\,\mathbf{h}}{\norm{\widehat{\mathbf{h}}}}\right|^2\right\rbrace$ incorporates this effect better than MSE $=\mathbb{E}\left\lbrace\norm{\mathbf{h}-\widehat{\mathbf{h}}}^2\right\rbrace$, where the impact of phase ambiguity on performance degradation diminishes with increasing SNR $\overline{\gamma}_{\rm id}$ values as shown in Figs.~\ref{fig:LSE-N} and~\ref{fig:LSE-1}, we preferred received power $p_r$ over MSE as metric to demonstrate the CE quality enhancement with increased $\overline{\gamma}_{\rm id}$.}
%

\begin{figure}[!t]
	\centering 
	\includegraphics[width=3.45in]{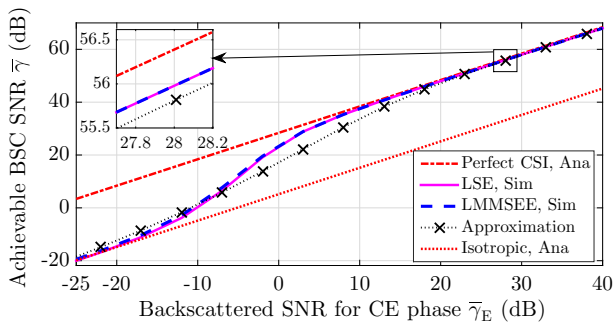}
	\caption{\small Validating the quality of the proposed approximation $\overline{\gamma}_{\rm a}$ for  $\overline{\gamma}$ available for ID with varying SNR $\overline{\gamma}_{\rm E}$ during CE for $K=N$. The average SNRs $\overline{\gamma}_{\rm id}$ and $\overline{\gamma}_{\rm is}$ for the two benchmarks are also plotted.}
	\label{fig:SNR_ana} 
\end{figure} 
\subsubsection{Tightness of Proposed Approximation for $\overline{\gamma}$}
Now we validate the quality of the closed-form approximation $\overline{\gamma}_{\rm a}$ proposed in Section~\ref{sec:apprG} for the average BSC SNR $\overline{\gamma}$ during ID phase. This result is important because $\overline{\gamma}_{\rm a}$ has been used for obtaining the joint optimal TA and PC by respectively exploiting the concavity and convexity of $\overline{\gamma}_{\rm a}$ in TA $\tau_c$ and integer constraint relaxed $K$.  So, we first consider $K=N$ and in Fig.~\ref{fig:SNR_ana} plot the analytical results for the backscattered SNR $\overline{\gamma}$ with (a) perfect CSI (as given by $\overline{\gamma}_{\rm id}$ in \eqref{eq:gammaID}), (b) LSE or LMMSEE (as given by $\overline{\gamma}_{\rm a}$ in \eqref{eq:GaM}), and (c) isotropic transmission (as given by $\overline{\gamma}_{\rm is}$ in \eqref{eq:gammaISO}). Whereas, the simulation results are plotted by averaging over the $10^5$ random channel realizations of LSE and LMMSEE based $\overline{\gamma},$ as respectively defined by $\overline{\gamma}_{\rm L}$ and $\overline{\gamma}_{\rm M}$ in Sections~\ref{sec:approxL} and~\ref{sec:approxM}. The validation results as plotted in Fig.~\ref{fig:SNR_ana} for varying BSC SNR $\overline{\gamma}_{\rm E}$ (defined in \eqref{eq:Kopt}) as available during the CE phase, show that for both low and high SNR $\overline{\gamma}_{\rm E}$  values    the  match between the analytical and simulation  is tight. This validates the quality of the proposed approximation $\overline{\gamma}_{\rm a}$ with a practically acceptable average gap between the analytical and simulation results of less than $1.7$dB in low CE SNR regime with $\overline{\gamma}_{\rm E}<-5$dB and less than $0.2$ dB  for the high SNR values $\overline{\gamma}_{\rm E}>15$dB available during CE phase. Thus, only in the range $-5\text{dB}<\overline{\gamma}_{\rm E}<15$dB, the match is not very tight. Further, the  $\overline{\gamma}_{\rm id}$ and  $\overline{\gamma}_{\rm is}$ plotted here again corroborate the earlier results in Figs.~\ref{fig:LSE-N} and~\ref{fig:LSE-1} that for the two extremes scenarios having very low $\overline{\gamma}_{\rm E}$ and very high $\overline{\gamma}_{\rm E}$, the average BSC SNR $\overline{\gamma}$ with LSE or LMMSEE respectively approaches the performance of isotropic transmission and as under  full beamforming gain with perfect CSI availability. 

\begin{figure}[!t]
	\centering 
	\includegraphics[width=3.48in]{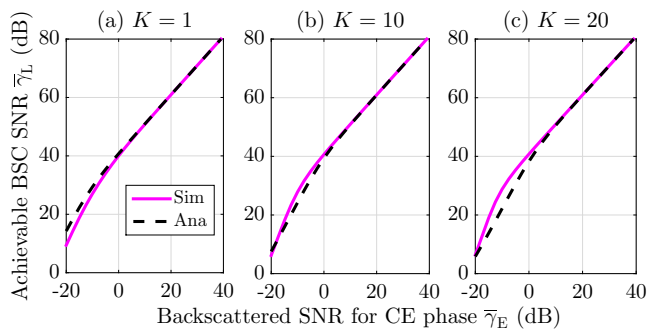}
	\caption{\small Verifying the tightness of the proposed approximation $\overline{\gamma}_{\rm a}$ for SNR $\overline{\gamma}_{\rm L}$ with LSE $\widehat{\mathbf{h}}_{\rm L}$ against varying PC $K$ for different $\overline{\gamma}_{\rm E}$ values.} 
	\label{fig:SNR_Ana_K} 
\end{figure} 
Lastly, we also verify that this approximation $\overline{\gamma}_{\rm a}$ for $\overline{\gamma}$ holds tight for varying PC $K$. For this, we plot the variation of analytical $\overline{\gamma}_{\rm a}$ and simulated values for $\overline{\gamma}_{\rm L}$ in Fig.~\ref{fig:SNR_Ana_K} with varying BSC SNR $\overline{\gamma}_{\rm E}$ values for different $K$ values. As believed, the analytical $\overline{\gamma}_{\rm a}$ provides a tighter match for the simulated $\overline{\gamma}_{\rm L}$ for $\overline{\gamma}_{\rm E}>0$dB. The average gap between the analytical $\overline{\gamma}_{\rm a}$ (cf. \eqref{eq:GaL} or \eqref{eq:GaM}) and simulated $\overline{\gamma}_{\rm L}$ results for $K=1$, $K=\frac{N}{2}=10$, and $K=N=20$ is respectively less than $0.06$dB, $0.09$dB, and $0.17$dB for $\overline{\gamma}_{\rm E}>5$dB. This completes the validation of the qualities of proposed LSE $\mathbf{h}_{\rm L}$, LMMSEE $\mathbf{h}_{\rm M}$, and the approximation $\overline{\gamma}_{\rm a}$. Next we use  these key analytical results for gaining the nontrivial design insights on joint optimal energy allocation and PC for CE at  $\mathcal{R}$.


\subsection{Insights on Optimal Design Parameters $\tau_{c_{\rm a}}$ and $K_{\rm opt}$}\label{sec:insights}
\begin{figure}[!t]
	\centering 
	\includegraphics[width=3.48in]{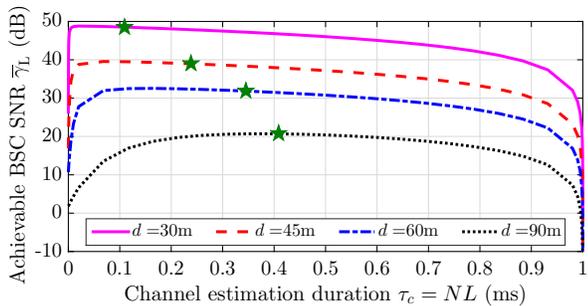}
	\caption{\small Validation of the unimodality of $\overline{\gamma}_{\rm L}$ in $\tau_c$ with $\tau=1$ms, $K=N$, and varying range $d$. The quality of approximation $\tau_{c{\rm{a}_N}}$ (marked as starred points)  for globally optimal $\tau_c$ is also verified.}
	\label{fig:tc_var_LSE} 
\end{figure} 
\subsubsection{Optimal TA $\tau_c$} Starting with an investigation on optimal TA $\tau_c$ for LS based CE with a given PC $K=N$ information, we first validate the claim made in Section~\ref{sec:quasi} regarding the quasiconcavity of $\overline{\gamma}$ (or $\overline{\gamma}_{\rm L}$ to be specific  in this case) in $\tau_c$. From Fig.~\ref{fig:tc_var_LSE}, where the variation of $\overline{\gamma}_{\rm L}$ (cf. \eqref{eq:gamma1}) with $\tau_c$ is plotted for different $\mathcal{R}$-to-$\mathcal{T}$ distance $d$ values, it can be   observed that $\overline{\gamma}_{\rm L}$ is quasiconcave or unimodal in TA variable $\tau_c$. Also, $\overline{\gamma}=\overline{\gamma}_{\rm L}=0$ for $\tau=\tau_c$, and the value of $\overline{\gamma}_{\rm L}$ at $\tau_c=0$  represents the performance under isotropic transmission. Further, we note that the proposed approximation $\tau_{c{\rm{a}_N}}$ (plotted as starred points in Fig.~\ref{fig:tc_var_LSE} and defined in Section~\ref{sec:approx-GOP}) provides a very tight match to the global optimal $\tau_c$, especially in high SNR regime (as represented by lower range $d$ values). Moreover, as for lower SNR scenarios,  more time needs to be allocated for accurate CE, $\tau_{c{\rm{a}_N}}$ is higher for   larger BSC range $d$ values. {Also, this investigation on optimal $\tau_c$, which is $<0.5\tau$ for practical SNR ranges, holds even for the high carrier frequency (in GHz range) applications with coherence time $\tau\approx100\mu$sec.}

\begin{figure}[!t]
	\centering 
	\includegraphics[width=3.4in]{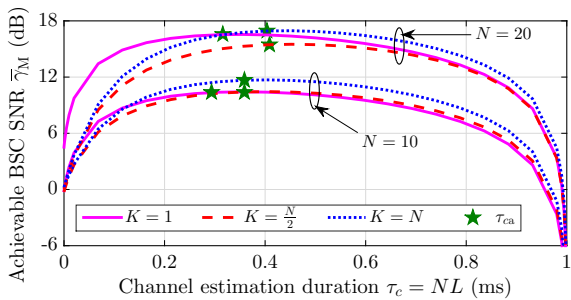}
	\caption{\small Variation of $\overline{\gamma}_{\rm M}$ with $\tau_c$ with $\tau=1$ ms, $d=100$m, and varying $K$. The quality of the approximation $\tau_{c{\rm{a}}}$ is also verified.}
	\label{fig:tc_var_LMMSE} 
\end{figure} 
Now we extend this investigation on optimal TA for a given PC by presenting the variation of average BSC SNR $\overline{\gamma}_{\rm M}$ for LMMSE based CE  with increasing number of antennas $N$ at $\mathcal{R}$ in Fig.~\ref{fig:tc_var_LMMSE} for different $K$ values. Again, we observe that, like  $\overline{\gamma}_{\rm L}$, $\overline{\gamma}_{\rm M}$ is quasiconcave in $\tau_c$. Moreover, $\tau_{c\rm a}$ closely approximates the optimal TA $\tau_c$ for CE that maximizes $\overline{\gamma}_{\rm M}$. This optimal TA $\tau_{c\rm a}$ increases for both higher $N$ and $K$ because more elements ($NK$ elements to be precise, from an $N\times K$ received signal matrix $\mathbf{Y}$) are required to be estimated using the same transmit power $p_t$. Also, it is noticed that the performance of LMMSEE with  $K=N$ and a relatively higher optimal TA $\tau_{c\rm a}$ has a better performance than that for $K=1$ with optimal TA. The latter holds because it enables to have a better quality estimate $\widehat{\mathbf{h}}_{\rm M}$ as obtained from a relatively larger sized  matrix $\mathbf{Y}\in\mathbb{C}^{N\times K}$ with sufficiently large CE time.

\begin{figure}[!t]
	\centering 
	\includegraphics[width=3.25in]{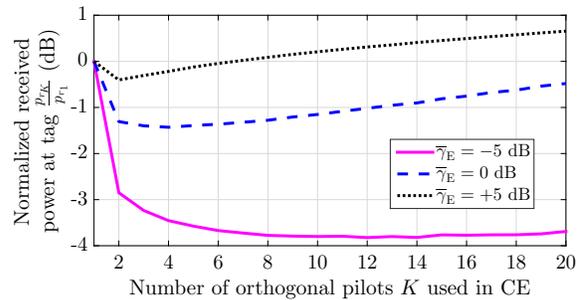}
	\caption{\small Variation of the received power $p_{r_K}$ at tag  with varying PC $K$ normalized to  power $p_{r_1}$ received with $K=1$ for different $\overline{\gamma}_{\rm E}$.}
	\label{fig:pilot_count_var} 
\end{figure} 
\subsubsection{Optimal PC $K$}\label{sec:res-OptK} For obtaining numerical insights on optimal PC $K=K_{\rm opt}$ for a given or fixed TA $\tau_c=\tau_{c_0}=0.1$ms as defined by \eqref{eq:Kopt}, in Fig.~\ref{fig:pilot_count_var} we plot the variation of the average  received power $p_{r}$ at $\mathcal{T}$ with PC $K$, denoted as $p_{r_K}$,  normalized to the power received with single pilot, denoted by  $p_{r_1}$, for varying $K$ and  $\overline{\gamma}_{\rm E}$. It can be clearly observed that the optimal PC $K_{\rm opt}$ is either $1$ or $N$, i.e., $K_{\rm opt}\notin\{2,3,\cdots,N-1\}$. Also, the average received power at $\mathcal{T}$ (like average backscattered SNR $\overline{\gamma}_{\rm a}$ for ID) is unimodal (but, convex) in $K$, implying that either of the two corner points will be yielding the maximum value of $p_r$. As with  $N=20$, $\overline{\gamma}_{\rm th}=3.32$dB, we notice that for $\overline{\gamma}_{\rm E}=-5$dB and $\overline{\gamma}_{\rm E}=0$dB, $K_{\rm opt}=1$, whereas for $\overline{\gamma}_{\rm E}=5$dB $>\overline{\gamma}_{\rm th}$, $K_{\rm opt}=N$. This validates the claims made in Section~\ref{sec:optK} and \eqref{eq:Kopt}. So, for low SNR $\overline{\gamma}_{\rm E}$ regime, when the propagation losses are severe during the CE phase, it is better to allocate all the transmit power $p_t$ to a single antenna and try to estimate an $N\times1$ vector $\mathbf{h}$ from a $N\times1$ received signal vector (cf. Section~\ref{sec:1P}) rather than distributing $p_t$ across $N$ antennas at $\mathcal{R}$ for estimating it from an $N\times N$ matrix (cf. Section~\ref{sec:NP}).

\begin{figure}[!t]
	\centering 
	\includegraphics[width=3.25in]{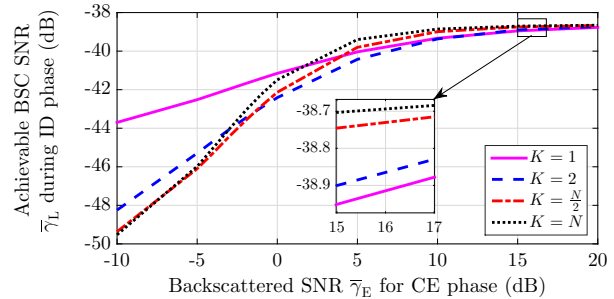}
	\caption{\small Variation of the average SNR $\overline{\gamma}_{\rm L}$ for ID using LSE $\widehat{\mathbf{h}}_{\rm L}$  with $\tau_c=\tau_{c_0}$, $K\in\{1,2,\frac{N}{2},N\}$ and  different SNR $\overline{\gamma}_{\rm E}$ values for CE.}
	\label{fig:Inisght_K_1_N_opt} 
\end{figure} 
To further corroborate the above mentioned claims, we plot the variation of the simulated backscattered SNR $\overline{\gamma}_{\rm L}$ during ID phase for varying $K$ and  $\overline{\gamma}_{\rm E}$ in Fig.~\ref{fig:Inisght_K_1_N_opt}. A similar result is obtained here showing that either $K=1$ or $K=N$ yields the best performance. Further, for lower $\overline{\gamma}_{\rm E}$, $K_{\rm opt}=1$ with $K=2$ performing better than both $K=\frac{N}{2}$ and $K=N$. Whereas as $\overline{\gamma}_{\rm E}$ increases and goes beyond $5$dB, $K_{\rm opt}=N$ and $K=\frac{N}{2}$ perform better than both $K=1$ and $K=2$.

\begin{figure}[!t]
	\centering 
	\includegraphics[width=3.48in]{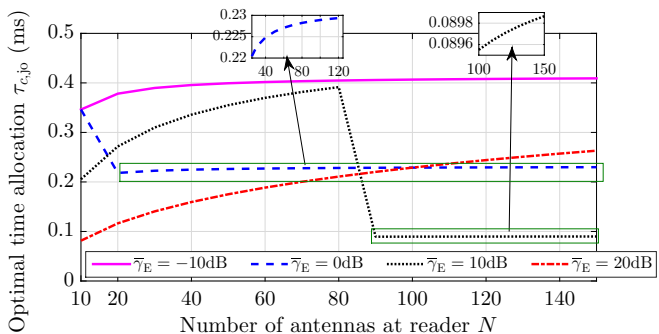}
	\caption{\small Insights on joint optimal TA $\tau_{c{\rm{a}}}$ and PC $K_{\rm opt}$ with increasing antennas $N$ at $\mathcal{R}$ for LS-based CE under different SNR $\overline{\gamma}_{\mathrm E}$ values.}
	\label{fig:OTA_OPC_Nvar} 
\end{figure} 
\subsubsection{Joint optimal TA and PC} Via Fig.~\ref{fig:OTA_OPC_Nvar} we finally present insights on the variation of joint optimal TA $\tau_{c,{\rm{jo}}}$ and PC $K_{\rm jo}$ as discoursed in Section~\ref{sec:joint} for LSE $\widehat{\mathbf{h}}_{\rm L}$ with increasing number of antennas $N$ at $\mathcal{R}$ under different SNR  $\overline{\gamma}_{\mathrm E}$ values during CE.  As $\overline{\gamma}_{\rm th}=\frac{\left(N-1\right)^2}{8\left(N+1\right)}$ in \eqref{eq:Kopt}  monotonically increases with  $N$, $K_{\rm jo}$ changes from $N$ to $1$ with increase in $N$. In particular, for low SNR $\overline{\gamma}_{\mathrm E}=-10$dB $=0.1$, $K_{\rm jo}=1,\,\forall\,N\in\{10,20,\cdots,150\}$ because $\overline{\gamma}_{\rm th}$ has a value of $0.92$ for $N=10$, which is higher than $0.1$. Due to similar reasons, $K_{\rm jo}=1$ for $N\le10$, $N\le82$, and $N\le802$ respectively with $\overline{\gamma}_{\mathrm E}=0$dB, $\overline{\gamma}_{\mathrm E}=10$dB, and $\overline{\gamma}_{\mathrm E}=20$dB. Otherwise, $K_{\rm jo}=N$. This can be observed from Fig.~\ref{fig:OTA_OPC_Nvar} in terms of the switching in  $\tau_{c,{\rm{jo}}}$  for $\overline{\gamma}_{\mathrm E}=0$dB and $\overline{\gamma}_{\mathrm E}=10$dB. Further, $\tau_{c,{\rm{jo}}}$, representing the tight approximation for optimal TA for CE, is higher for lower  $\overline{\gamma}_{\mathrm E}$ to have more time for CE enabling a better quality LSE for $\mathbf{h}$. Further, for both $K_{\rm jo}=1$ and $K_{\rm jo}=N$, $\tau_{c,{\rm{jo}}}$ increases with increasing $N$ as more elements need to be estimated using the same  training power $p_t$. Owing to the same need,  $\tau_{c,{\rm{jo}}}$ is higher for $K_{\rm jo}=N$ as compared to that for $K_{\rm jo}=1,\,\forall N>1$.


\subsection{Performance Gain and Comparisons}\label{sec:comp}
\begin{figure}[!t]
	\centering 
	\includegraphics[width=3.48in]{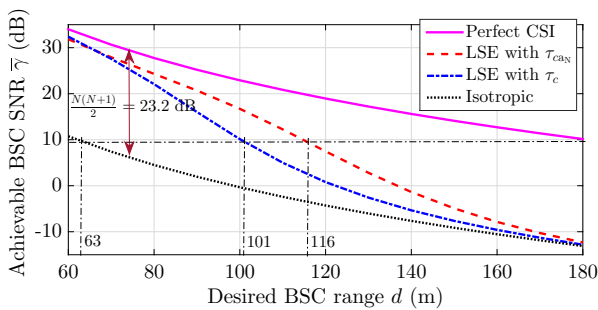}
	\caption{\small Variation of $\overline{\gamma}$ for the proposed  LSE $\widehat{\mathbf{h}}_{\rm L}$  with $\tau_c=\tau_{c_0}$ and  $\tau_c=\tau_{c{\rm{a}_N}}$. The resulting enhancement in BSC range $d$ for $K=N=20$ while satisfying  SNR requirement of $\overline{\gamma}=10$ dB  is also shown.}
	\label{fig:comp_LSE} 
\end{figure} 
In this part of the results section, we quantify the enhancement in $\overline{\gamma}$ as achieved by optimizing the
TA $\tau_c$ and PC $K$ for efficient CE. Specifically, in Fig.~\ref{fig:comp_LSE} we plot the variation of the achievable $\overline{\gamma}_{\rm L}$ with LSE $\widehat{\mathbf{h}}_{\rm L}$ and optimal TA $\tau_c=\tau_{c{\rm{a}}}$ for fixed PC $K=N$ and different BSC ranges $d$. The variations of $\overline{\gamma}$ for  isotropic radiation and directional transmission with perfect CSI are also plotted along with the BSC using LSE $\widehat{\mathbf{h}}_{\rm L}$ with fixed $\tau_c=\tau_{c_0}=0.1$ms for comparison. The efficacy of using an antenna array at $\mathcal{R}$ can be observed from the fact that the BSC range $d$ gets enhanced from {$63$m to $101$m} by using the proposed LSE  $\widehat{\mathbf{h}}_{\rm L}$ based precoder and combiner designs at $\mathcal{R}$ with $N=20$ for achieving $10$ dB backscattered SNR in comparison to the isotropic transmission. Further, if instead of fixed $\tau_c=\tau_{c_0}$, optimized time allocation $\tau_c=\tau_{c{\rm{a}}}$ is considered for designing the CE and ID phases, then this improvement in BSC range increases to {$116$m}. Overall, the proposed optimal time (or energy for fixed $p_t$) allocation $\tau_c=\tau_{c{\rm{a}}}$  yields an average improvement of $3$ dB (two-fold gain) in the achievable $\overline{\gamma}_{\rm L}$ with fixed TA $\tau_c=\tau_{c_0}$ for the CE phase with varying BSC range $d$ from {$60$m to $180$m}.

\begin{figure}[!t]
	\centering 
	\includegraphics[width=3.48in]{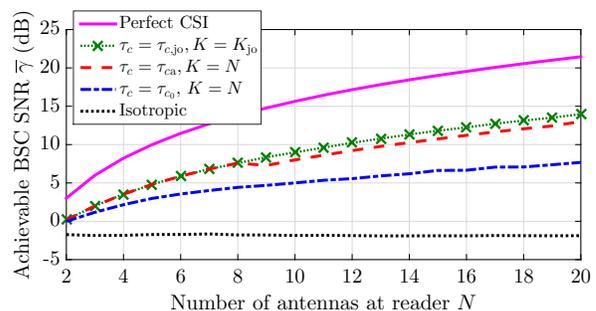} 
	\caption{\small Variation of $\overline{\gamma}_{\rm M}$ for LMMSEE $\widehat{\mathbf{h}}_{\rm M}$  with $N$ at $\mathcal{R}$ for (a) $\tau_c=\tau_{c_0},K=N$, (b) $\tau_c=\tau_{c{\rm{a}}},K=N$, and (c)  $\tau_c=\tau_{c,{\rm{jo}}},K=K_{\rm jo}$.}
	\label{fig:comp_LMMSE} 
\end{figure} 
Next we extend this result for LSE to a similar comparison study, but now with LMMSE based CE. In particular, in Fig.~\ref{fig:comp_LMMSE} we   compare the achievable BSC SNR $\overline{\gamma}_{\rm M}$ performance of LMMSEE $\widehat{\mathbf{h}}_{\rm{M}}$ with joint optimal TA $\tau_{c,\rm jo}$ and PC $K_{\rm jo}$ against that of $\widehat{\mathbf{h}}_{\rm{M}}$ for optimal TA $\tau_{c\rm a}$ with $K=N$ and $\widehat{\mathbf{h}}_{\rm{M}}$ with fixed TA $\tau_c=\tau_{c_0}=0.1$ms and $K=N$. Again, here the benchmark perfect CSI and isotropic transmission cases are also plotted. From Fig.~\ref{fig:comp_LMMSE} it can be observed that there is no gain achieved by joint optimal TA and PC over optimal TA alone with $K=N$ for $N\le8$, because the underlying $\overline{\gamma}_{\rm th}<\overline{\gamma}_{\rm E_1}$. However, for $N>8$ as $K_{\rm jo}=1$,  $\overline{\gamma}_{\rm M}$ with $K=K_{\rm jo}=1$ and $\tau_c=\tau_{c,\rm jo}$ yields improvement over that with $\tau_c=\tau_{c\rm a}$ and fixed PC $K=N$. The average improvement provided by LMMSEE with fixed $\tau_c=\tau_{c_0}$ and PC $K=N$ is about $4.3$dB in terms of $\overline{\gamma}_{\rm M}$ over the isotopic transmission for different values of $N$ ranging from $2$ to $20$. Further, optimal TA $\tau_c=\tau_{c\rm a}$ with fixed PC $K=N$ can provide an improvement of $6.7$dB over fixed TA $\tau_c=\tau_{c_0}$. Moreover, the joint optimal TA  $\tau_c=\tau_{c,\rm jo}$ and PC $K=K_{\rm jo}$  provides an additional average improvement of about  $2.6$dB over optimal TA $\tau_c=\tau_{c\rm a}$ with fixed PC $K=N$.

\begin{figure}[!t]
	\centering 
	\includegraphics[width=3.48in]{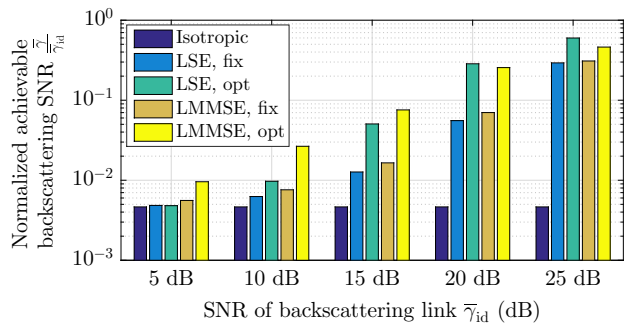}
	\caption{\small Comparison of various CE schemes in terms of the average BSC SNR $\overline{\gamma}$ as normalized to the one under perfect CSI availability.}
	\label{fig:hist_SNR} 
\end{figure} 
Lastly, we corroborate the utility of the proposed analysis and optimization by quantifying the underlying achievable gains in terms of the average BSC SNR $\overline{\gamma}$. Specifically, via Fig.~\ref{fig:hist_SNR} the achievable SNR $\overline{\gamma}$ as normalized to the maximum value $\overline{\gamma}_{\rm id}$ achieved with perfect CSI availability for different schemes is compared. Apart from the isotropic transmission having average BSC SNR $\overline{\gamma}_{\rm is}$, two fixed benchmark schemes, namely, LSE and LMMSEE with fixed TA $\tau_c=\tau_{c_0}$ and PC $K=N$ are compared against the proposed LSE and LMMSEE with jointly optimized TA $\tau_{c,\rm jo}$ and PC $K_{\rm jo}$. With increasing $\overline{\gamma}_{\rm id}$, implying better channel conditions, the achievable BSC SNR for each scheme, except the isotropic transmission, increases due to the underlying enhancement in the CE quality and approaches the value  $\overline{\gamma}_{\rm id}$ achieved with perfect CSI availability for $\overline{\gamma}_{\rm id}>30$dB. For the isotropic transmission, the normalized SNR $\frac{ \overline{\gamma}_{\rm is}}{\overline{\gamma}_{\rm id}}=\frac{2}{N\left(N+1\right)}=0.0047$ is independent of $\overline{\gamma}_{\rm id}$ because there is no CE involved. The LSE and LMMSEE with fixed TA and PC respectively provide  $57$ and $58$ times more BSC SNR for ID as compared to that  with isotropic transmission. Here, the LMMSEE based $\overline{\gamma}$, as denoted by $\overline{\gamma}_{\rm M}$, respectively provides about $14.1\%$ and $41.3\%$ over its LSE counterpart $\overline{\gamma}_{\rm L}$, with and without joint optimization. The gains achieved by the joint optimization  over the fixed TA and PC for LSE and LMMSEE are about $2.5$dB and $3$dB respectively. But, for high SNR regime, LSE can perform as good as LMMSEE both with and without optimal TA-PC. The joint optimization is actually very important  in the practical SNR regime of $10$dB to $25$dB (cf. Fig.~\ref{fig:hist_SNR}). Hence, for low SNR scenarios, LMMSEE with joint optimal TA-PC should be preferred. Whereas, for high SNR applications, LSE with $\tau_c=\tau_{c,\rm jo}$ and $K=K_{\rm jo}$ can be adopted to avoid complexity overhead or the need for prior information on $\mathbf{C}_{\mathbf{h}_{\rm v}}$.

\section{Concluding Remarks}\label{sec:concl} 
We presented a novel joint CE, energy and pilot count allocation investigation for a full-duplex monostatic BSC setup with a  multiantenna reader $\mathcal{R}$. We first obtained a robust channel estimate yielding the global LS minimizer while satisfying a rank-one constraint on the backscattered channel matrix. Using the proposed principal eigenvector approximation for the equivalent real domain transformation of the LS problem, the LMMSEE for the BSC channel is obtained while accounting for the impact of orthogonal PC used during the CE phase. These LSE and LMMSEE are used to design a MRT precoder and MRC combiner at $\mathcal{R}$ during the ID phase. Then exploring the concavity of the tight approximation of average SNR $\overline{\gamma}$ in $\tau_c$ for a fixed transmit power $p_t$ and convexity in integer relaxed PC $K$, it was shown that the protocol designed using the jointly optimized TA $\tau_c=\tau_{c,\rm jo}$  and PC $K=K_{\rm jo}$ nearly \textit{doubles} the achievable performance with fixed TA $\tau_c=\tau_{c_0}$ and PC $K=N$. It was also proved that the optimal PC is either given by $K=1$ or $K=N$. Further, we showed that LSE and LMMSEE with optimized TA and PC should be respectively deployed for the high and low SNR regimes. Thus, this work corroborates the significance of the joint optimal CE and resource allocation between CE and ID phases for maximizing the efficacy of the antenna array at $\mathcal{R}$ in realizing long range QoS-aware BSC from a passive tag. \textcolor{black}{In future we would like to extend this investigation to design the optimal training sequences for multi-tag MIMO BSC systems.}

\bibliographystyle{IEEEtran}
\bibliography{refs-CSI-BSC_TSP}

\end{document}